# Exoplanet Biosignatures: Future Directions

Sara I. Walker[1,2,3,4*], William Bains[5], Leroy Cronin[7], Shiladitya DasSarma[8], Sebastian Danielache[9,10], Shawn Domagal-Goldman[11,12], Betul Kacar[13,14,15], Nancy Y. Kiang[15,16], Adrian Lenardic[17], Christopher T. Reinhard[18,19], William Moore[19,20], Edward W. Schwieterman[4,12,18,21,22], Evgenya L. Shkolnik[1], Harrison B. Smith[1]

[1] School of Earth and Space Exploration, Arizona State University, Tempe AZ USA

[2] Beyond Center for Fundamental Concepts in Science, Arizona State University, Tempe AZ USA

[3] ASU-Santa Fe Institute Center for Biosocial Complex Systems, Arizona State University, Tempe AZ USA

[4] Blue Marble Space Institute of Science, Seattle WA USA

[5] EAPS (Earth, Atmospheric and Planetary Sciences), MIT, 77 Mass Ave, Cambridge, MA 02139, USA

[6] Rufus Scientific Ltd., 37 The Moor, Melbourn, Royston, Herts SG8 6ED, UK

[7] School of Chemistry, University of Glasgow, Glasgow, G12 8QQ UK

[8] Department of Microbiology and Immunology, Institute of Marine and Environmental Technology, University of Maryland School of Medicine, Baltimore, MD, USA

[9] Department of Materials and Life Science, Faculty of Science and Technology, Sophia University, Tokyo, Japan.

[10] Earth Life Institute, Tokyo Institute of Technology, Tokyo Japan.

[11] NASA Goddard Space Flight Center, Greenbelt, MD, USA

[12] NASA Astrobiology Institute, Virtual Planetary Laboratory Team, University of Washington, Seattle, WA, USA

[13] Organismic and Evolutionary Biology, Harvard University, Cambridge, MA, USA

[14] NASA Astrobiology Institute, Reliving the Past Team, University of Montana, Missoula, MT, USA

[15] University of Arizona, Department of Molecular and Cell Biology and Astronomy, Tucson, AZ

[15] NASA Goddard Institute for Space Studies, New York, NY USA.

[16] Department of Earth Science, Rice University, Houston, TX, USA

[17] School of Earth and Atmospheric Sciences, Georgia Institute of Technology, Atlanta, GA, USA

[18] NASA Astrobiology Institute, Alternative Earths Team, University of California, Riverside, CA, USA

[19] Department of Atmospheric and Planetary Sciences, Hampton University, Hampton, VA, USA

[20] National Institute of Aerospace, Hampton, VA, USA

[21] Department of Earth Sciences, University of California, Riverside, CA, USA

[22] NASA Postdoctoral Program, Universities Space Research Association, Columbia, MD, USA

*Author for correspondence. Email: sara.i.walker@asu.edu Address: School of Earth and Space Exploration,

Arizona State University, PO Box 871404 Tempe AZ USA 85287-1404. Phone: 480.727.2394

Authors are listed alphabetically with the exception of the first author.

**Abstract:** Exoplanet science promises a continued rapid accumulation of new observations in the near future, energizing a drive to understand and interpret the forthcoming wealth of data to identify signs of life beyond our Solar System. The large statistics of exoplanet samples, combined with the ambiguity of our understanding of universal properties of life and its signatures, necessitate a quantitative framework for biosignature assessment Here, we introduce a Bayesian framework for guiding future directions in life detection, which permits the possibility of generalizing our search strategy beyond biosignatures of known life. The Bayesian methodology provides a language to define quantitatively the conditional probabilities and confidence levels of future life detection and, importantly, may constrain the prior probability of life *with or without* positive detection. We describe empirical and theoretical work necessary to place constraints on the relevant likelihoods, including those emerging from stellar and planetary context, the contingencies of evolutionary history and the universalities of physics and chemistry. We discuss how the Bayesian framework can guide our search strategies, including determining observational wavelengths or deciding between targeted searches or larger, lower resolution surveys. Our goal is to provide a quantitative framework not entrained to specific definitions of life or its signatures, which integrates the diverse disciplinary perspectives necessary to confidently detect alien life.

# Table of Contents

# 1   Introduction

Over the last two decades, hundreds of exoplanets have been discovered orbiting other stars, inspiring a quest to understand the diversity of planetary environments that could potentially host life. Soon we will be positioned to search for signs of life on these worlds. Upcoming missions are targeted at obtaining atmospheric spectra of planets that may, like Earth, sustain liquid water oceans on their surfaces. To date, efforts to identify biosignatures on alien worlds have focused on the dominant chemical products and surface features of examples of life known from Earth, as well as some theoretically modeled cases (reviewed by Schwieterman *et al*., 2017, this issue). If we are lucky, we may be able to identify "Earth-like" life on "Earth-like" worlds. If we are unlucky, and true Earths with Earth-like life are rare, our current approaches could entirely fail to discover alien life or to place constraints on the processes of life or their frequency. Expanded efforts are necessary to develop *quantitative* approaches to remote biosignature detection, applicable both in cases where the stellar or planetary context, or biochemistry is like Earth, and in cases where these diverge significantly from what is known from Earth.

With the exception of modern Earth, there are currently no known planets that can provide an unambiguous, easily detectable, true-positive biosignature of life – a so-called "smoking gun". Even Earth throughout most of its history may not have had remotely detectable biosignatures, despite the presence of life (Reinhard *et al*., 2017). A major challenge is that the diversity of exoplanets greatly exceeds the variety of planetary environments found within our own solar system, such that the majority of exoplanets have no analogs in our solar system. Examples include water worlds, massive rocky planets, and small ice giants. The majority of discovered exoplanets orbit low-mass stars, and are subjected to very different radiation and space plasma environments than planets in our own solar system (Coughlin *et al*., 2016; Twicken *et al*., 2016). A metabolic product such as $O_2$ might be a smoking gun signal of life on one world and not a biosignature at all on another, leading to the possibility of false positives (see Meadows *et al*., 2017, this issue, for an in-depth treatment of $O_2$). Given the limited data we can collect on exoplanets (see review of observation capabilities in Fujii *et al*., 2017, this issue), and the stochasticity of planetary evolution (Lenardic, *et al*., 2012), we may only be able to predict the properties of exoplanets statistically (Iyer *et al.,* 2016; Wolfgang *et. al.,* 2016). Our uncertainty in exoplanet properties such as bulk composition, geochemistry, and climate – due both to lack of knowledge, and technical limitations on what we can directly infer from observational data – is a major hurdle to be overcome in our search for life outside our own solar system.

A major hurdle is that we face signficant uncertainty in our understanding of what life is (Chyba and Cleland, 2002; Walker and Davies, 2013). Our views of life and its defining features have expanded in recent years with discoveries of novel metabolisms (Sogin *et al*., 2006; Rappe *et al*., 2003; Hughes *et al*., 2001; Li and Chen 2015), and advances in synthetic biology and systems chemistry, which challenge our assumptions about what chemistries can participate in terrestrial life and in prebiotic chemistry (Chaput *et al.,* 2012; Malyshev *et al.,* 2014; Sadownik, 2016). From a first principles perspective, life is more readily understood in terms of *dynamic processes* than *chemical products*. Yet, in biosignature research for exoplanets we so far have focused primarily on the chemical products of Earth's life. The focus on chemical products has primarily been driven by practical limitations of current detection methods: current or planned exoplanet missions will be geared to detect the presence or absence of materials, leading to a focus on what materials could be biologically derived. In particular, this led to a focus on the idea of a smoking gun biosignature. $O_2$ is the most notable example; however, even beyond the challenges associated with false positives, this particular smoking gun was not abundant in Earth's atmosphere for several billion years of its history (Lyons *et al*., 2014), rendering Earth's life undetectable by current methods for the first few billion years. Thus, even though the *process* of photosynthesis was present, our current product-based strategy would miss detecting it on inhabited worlds like early Earth. Ultimately, in our search to discover life, we are interested in answers to questions like: how frequently photosynthesis (or other life processes) evolve in a given planetary context? Beyond the biosignature community, life is not typically characterized in terms of its products, but instead in terms of its processes. Hallmark features of life, such as information processing, metabolism, reproduction, homeostasis, evolution *etc.* are all *processes*, which may generate different products in different evolutionary and environmental contexts. To advance our capabilities for life detection, next generation biosignature research must bridge our product-based detection strategy with an understanding of the underling living processes, in order to identify signatures of life in diverse planetary contexts.

A process-based understanding will allow extrapolation to contexts different from Earth, where presumably the same universal processes of life (e.g., evolution, information processing, metabolism) should operate, but may lead to very different outcomes – that is to different remotely detectable products of life.  The multitude of exoplanets discovered provides unprecedented opportunity to address fundamental questions regarding the nature and *distribution* of life with large statistical data sets, but we must first better understand the processes governing both planets and life. Bridging processes with detectable products necessitates new cross-disciplinary collaborations. To make progress we must address the

questions:

- *What fundamental life processes could underlie the chemistry we can detect on exoplanets?*

- *How do we infer the presence (or absence) of these processes?*

- *How can understanding the processes of life inform new ways to identify and interpret the chemical signatures of life?*

In this manuscript, we introduce a Bayesian framework appropriate and timely for the long-term goal of searching exoplanets for signs of life. A Bayesian methodology provides a language to define quantitatively the conditional probabilities and confidence levels of future life detection and, importantly, may constrain the prior probability of life *with or without* positive detection. To understand what is needed to quantify these probabilities, we review emerging and future developments of the study of life processes, their origins, their planetary contexts, the integrated tools necessary to model them, and the methodological tools necessary to detect their consequences. These inevitably require continued expansion of a cross-disciplinary community to develop the conceptual frameworks required to interpret the increasing (yet sparse) data upon which claims for the presence of life beyond our solar system will eventually be made. Because we are focused on future directions, we note that the views presented herein do not represent community consensus. The myriad challenges that come with adopting a probabilistic framework for life detection drive the organization of this paper, as resolving these challenges should be a priority for the exoplanet research community over the coming decade.

## 2 Setting the Stage: What is life? What is a biosignature?

Des Marais *et al.*, (2002) defined a *biosignature* as an "object, substance, and/or pattern whose origin specifically requires a biological agent" (Des Marais and Walter, 1999; Des Marais *et al*., 2008). In this paper we follow this convention and refer to a substance or pattern that is known to be an indicator of biological activity (in a given planetary context) as a biosignature, e.g., a 'biosignature molecule' or 'biosignature pattern.' More specifically, we quantify a ***biosignature as a molecule, pattern or other signal that has a non-zero probability of occurring, conditioned on the presence of a living process*** (see Section 3 where we define *P(data|life)* and provide a quantitative definition for a biosignature within a Bayesian framework). Importantly, a biosignature does not imply life, it only implies a signal consistent with life. To qualify as *evidence* for life, a biosignature should be much more likely to be produced by living processes than by abiotic ones (see Section 3 for an in-depth discussion, and Catling *et al*., 2017, this issue, for additional perspective). That is, a molecule, pattern or signal must be able to be produced by life to be a biosignature, but it does not qualify as evidence for life unless life is the best explanation for its production in a given environmental context.

A challenge for developing a quantitative framework for assessing biosignature candidates is that life – the very thing we hope to measure – is notoriously difficult to define. For example, the definition of Des Marais *et al.* (2002) specifically refers to biological agency, yet we are far from a quantitative framework that precisely captures what we mean by "agent" (Barandiaran *et al*., 2009). The state of the field is such that more than 100 definitions for life exist, alongside many attempts to analyze them (Chyba and McDonald 1995; Kolb 2007; Benner 2010;Trifonov, 2011; Mix, 2015). Some of the most common words used in defining life are shown in Table 1, demonstrating just how far from consensus we truly are. Some have argued that it does not make sense to define life until we have a theory for life (Cleland, 2012; Walker, 2017), much in the same way water was only precisely defined as $H_2O$ after the advent of molecular theory (Cleland and Chyba, 2002). A thorough review on the literature of attempts to define life is outside of the scope of this paper; however it is important to acknowledge the challenges we face in biosignature research due to ambiguities in our ability to precisely quantify what is 'life' is.

*Table 1. Most common words used in definitions for life, from Trifonov (2011).*

| life | organic | internal | capacity |
|------|---------|----------|----------|
| living | alive | replication | different |
| system | evolution | being | force |
| matter | materials | change | form |
| systems | reproduction | characteristics | functional |
| environment | existence | entity | highly |
| energy | defined | external | . |
| chemical | growth | means | mutation |
| process | information | molecules | necessary |
| metabolism | open | one | network |
| organisms | processes | order | objects |
| organization | properties | organisms | only |
| complexity | property | state | organized |
| ability | reproduce | things | reactions |
| itself | through | time | self-reproduction |
| able | complex | way | some |
| capable | evolve | based | three |
| definition | genetic | biological | |

Due to our lack of quantitative understanding of life, standards for the search for life have historically been *qualitative* in nature (true for both exoplanets and within the Solar System). As an example, consider an approach to the search for life that follows the adage *"I'll know it when I see it."* For different disciplines *"it"* means different things. For example, biochemists might cite the molecular species that constitute 'life-as-we-know-it', such as DNA, RNA and amino acids, whereas, a physicist might discuss the emergence of collective behavior and so on. We outline some of these differences between disciplines, based on our own experiences as researchers in diverse areas in Table 2. The table is not intended to be exhaustive (which would be a research program in its own right), nor representative of a majority opinion, but merely to highlight how diverse, and controversial, approaches to the question "what is life?" can be.

In order to evolve into a scientific discipline with testable hypotheses, biosignature science needs to make *quantitative* predictions based on the hypothesis that life is or is not present in a given environment. Gradually, we are developing a language and the quantitative frameworks required for this, but further progress will require even greater convergence of the disciplinary perspectives in Table 2 and mo.  No one discipline is "right" with respect to its perspectives on life, and each paints just one part of an emerging picture of what could be the most universal and fundamental properties of life. In what follows, we leverage this diversity of perspectives, unifying them within a common Bayesian formalism for searching for life on other worlds. The goal is to liberate our search strategies as much as possible from being entrained to specific definitions for life or its signatures, and instead to frame the problem in terms of what is observable and what we can infer from those observables based on what is known about non-living and living processes. As this paper illustrates, we have much work ahead as a community to realize the promising future directions that could finally enable us to detect life on another world (and be confident in our assertion of success).

*Table 2. Disciplinary perspectives on signatures of living processes.*

| Scientific discipline | Typical measures of life and objects of study | Biosignature relevance |
|---|---|---|
| Mathematics | Theorems, proofs, calculus, algebra, number theory, geometry, probability and statistics, computational science | The language of science. Quantitive frameworks of relationships in nature. |
| Physics | Motion of mass and electromagnetic Energy, quantum behavior, organization, dissipative structure, collective behavior, emergence, information, networks, molecular machines | Conservation laws to constrain abiotic context. Systems interactions of biological processes |
| Chemistry, Biophysics | redox potential, Gibbs free energy (Hoehler *et al.*, 2007) | Metabolic processes that alter the redox state of the environment |
| Microbiology, Molecular Biology, Biochemistry | Cells, genes, genomes, proteins, metabolism | Constraints on evolutionary path requirements for a type of life to emerge. Metabolic products that can be strictly biogenic. |
| Geologists, Geophysics | Isotope fractionation, morphology, fossils | Planet formation factors that determine prebiotic elements. Plate tectonics to allow a carbon cycle. |
| Philosophy | Emergence, meaning, goal-directedness | Definitions of intelligence, optimality. |
| Ecology | Ecosystem, community dynamics, scaling laws, keystone species (May, 1974; May and Saunders, 2007; Pikuta *et al.*, 2007; Amaral-Zettler *et al.*, 2011) | System interactions that lead to dominance or community mixes of particular kinds of life, determining what biosignatures will be detectable. |

| Biochemistry, Geochemistry | Elemental cycling (Schlesinger, 2013), serpentinization | Budgeting of the fluxes and stocks of particular molecules, wherein the net accumulated stock or phasing of fluxes may be detectable biosignatures. |
| --- | --- | --- |
| Astronomy | planetary-scale spectral signatures, molecular line lists, remote observation (Seager 2014; Seager *et al*., 2016; Meadows *et al*., 2005) | Stellar context for life determines the radiative balance and elemental composition of a planet. Detection of biosignatures in planetary spectra from transits or direct imaging. |

# 3   Detecting unknown biology on unknown worlds: A Bayesian Framework

To qualify as evidence for life in a given environment, a biosignature should be much more likely to be produced by living processes than by abiotic ones. For example, with some caveats (Meadows *et al*., 2017, this issue) current understanding provides confidence that geochemistry on a planet bearing liquid water will not generate an atmosphere containing >1% $O_2$, so $O_2$ is *a priori* a good biosignature. However, $O_2$ as a biosignature may be rare: the likelihood of oxygenic photosynthesis on other worlds is unknown. What is known is that for 85% of Earth's history, life did not produce significant amounts of atmospheric $O_2$. By contrast, we might expect that if life exists on a world with hydrothermal systems and sulfate in its oceans, life will evolve to produce $H_2S$; however, we are also confident that hydrothermal systems on such a world will make $H_2S$ abiotically, too. So, $H_2S$ on such a world would be an ambiguous indicator of life.

**Terminology**

*Biosignature*: an object, substance, and/or pattern of biological origin, such that $P(data|life) > 0$
*Detectability:* confidence of biological origins for an observed biosignature signal, D = P(data|life)/P(data|abiotic) (in the absence of noise). A biosignature is an indicator of life if D > 1.
*Habitability:* conditions suitable for life, it is commonly implied that $P(life) > 0$ for 'Earth-like' life
*False positive:* abiotic observables that mimic biologically produced observables, where $P(data|life) > 0$ and $P(data|abiotic) > 0$
*False negative*: biosignatures that are not detectable, with $P(data|life) \sim 0$, despite the presence of life.
*Anti-biosignature:* an object, substance, and/or pattern such that $P(data|life) = 0$

These examples illustrate how, to claim detection of life, measurements must be statistically quantified within the context of our expectations. Here, we introduce a Bayesian framework for guiding future directions in life detection, which permits the possibility of generalizing our search strategy beyond biosignatures of known life. We incorporate process-based approaches to constrain the probabilities of both living and nonliving processes to generate a particular observational signal, as required for Bayesian inference. Catling *et al*. (2017 this issue) suggest evaluation of four sets of criteria in order: (1) the stellar properties of the exoplanetary system (for example, if the planet can support surface liquid water), (2) characterization of the exoplanet surface and atmosphere, (3) identification of biosignatures in the available data and (4) exclusion of false positives. Their proposed scheme is based on current knowledge of biosignatures to increase confidence levels within a Bayesian framework, and is based on production of biosignatures similar to those of known life.  Here, we focus on unifying diverse research areas within a common quantitative framework to better constrain likelihoods of living and non-living processes, providing a means to organize current and future data in the assessment of upcoming observational data.

Bayesian inference permits evaluating the probability of a hypothesis (*e.g.,* the presence of life) given a set of observed data. The *posterior probability* quantifies the probability of a hypothesis once the evidence has been taken into account. It is calculated based on *prior probability*, quantifying the probability a hypothesis is true, and a *likelihood function,* which quantifies the compatibility of the evidence with the hypothesis, that is, the probability of observing the data given the hypothesis. Specifically, a Bayesian claim of detection of life requires quantifying:

- The likelihood of the signal arising due to living processes.

- The likelihood of the signal arising due to abiotic processes.

- The prior probability of the living process.

These likelihoods are cast in terms of conditional probabilities, where a *conditional probability* is the likelihood of observing an event, given another event has already occurred. For example, the conditional probabilities $P(H_2S|anaerobic\ respiration)$ and $P(H_2S|hydrothermal\ systems)$ quantify the likelihood of abundant atmospheric $H_2S$ arising due to living processes or to abiotic processes, respectively (here and throughout the "|" operator means "given" or "conditioned on" and indicates a conditional probability). $H_2S$ is not a good biosignature in the example provided earlier precisely because biotic and abiotic production are both potentially important, such that $P(H_2S|anaerobic\ respiration)$ ~ $P(H_2S|hydrothermal\ systems)$ without additional contextual information. Likewise, on modern Earth $O_2$ is a good biosignature because the likelihood of it arising due to life, $P(O_2|\ oxygenic\ photosynthesis)$, is much higher than by the abiotic processes of photodissociation or volcanic outgassing, quantified as $P(O_2|abiotic)$, e.g. we expect $P(O_2|\ oxygenic\ photosynthesis) >> P(O_2|abiotic)$.

In modeling biosignatures we have so far focused on the likelihood of generating a particular set of observational signatures, given the presence of life. However, soon we will have observational data to actually search for life. In analyzing this data we are interested in the inverse problem: ***what is the likelihood of life, given a set of observational data?***

A Bayesian framework permits determining the *posterior probability of life* (e.g., the likelihood of life, given observational data), for a given set of observational data, with the basic conditional probability:

(1) $P(life|data) = \frac{P(data|life)P(life)}{P(data)}$

Here, data is intended to indicate any observable that be indicative of life, such as NIR absorption features in the case of exoplanets. The denominator is the total likelihood of observing a given data set, and can be expanded further:

(2)

$$P(life|data) = \frac{P(data|life)P(life)}{P(data|\neg life)(1 - P(life)) + P(data|life)P(life)}$$

The "$\neg$" logical operator means "not", where $P(data|\neg life)$ is the probability of the data in the absence of life (given no life), and $P(\neg life) = 1 - P(life)$ is the probability there is no life.

*P(life|data)* is what we would like to know: the posterior probability of life, given a set of observational data. To determine the likelihood of life in a given data set, we must tightly

constrain *P(data|life)* the probability of the observational data given life is present, and $P(data|\neg life)$, the probability of the observations arising if life is not present. This latter term includes contributions from abiotic sources (life is absent) or experimental noise:

*( 3 )* $P(data|\neg life) = P(data|abiotic) + P(data|noise)$

Additionally, knowledge of *P(life),* the prior probability of living processes, is required to assess the likelihood of life. In Catling *et al.* (2017, this issue), the conditional probabilities include an explicit term for *context* in terms of joint probabilities, *e.g., P(data|life)* was instead cast as *P(data|life, context)*. Here, we include context as implicit in the probabilities for living or nonliving processes as these should in any case be conditioned on what is known about the context for the observation (and the abiotic probability could by definition be considered the 'context'). In general, we care about the likelihood of life given a particular context, and not necessarily the probability of the context. We discuss the importance of context and how those terms arise in a Bayesian framework more explicitly in Section 7 on *P(life),* below.

The utility of the Bayesian approach is that it permits separating the calculation of the prior probability of life, *P(life),* from the likelihood of observational data if life is present *P(data|life)* or if life is not present *P(data|abiotic)*. That is, it permits quantifying the ***detectability*** of life, and thereby provides a tool for identifying promising targets in our search for life, without necessarily knowing the prior probability of life itself, which is currently unconstrained (discussed more below in Section 6). In the Bayesian framework, detectability can be quantified as:

$$( 4 )\ D = \frac{P(data|life)}{P(data\,|\,abiotic)+P(data\,|\,noise)}$$

where the denominator is again, the probability that the signal was not generated by living processes. In the limit there is no experimental noise:

$$( 5 )\ D_{noise\to0} = \frac{P(data|life)}{P(data|abiotic)}$$

Comparing the likelihood of the data being produced by life to its likelihood to be produced by abiotic sources can provide a guide to how likely we are to detect life on a given target, under the assumption that life exists on that target. In other words, detectability provides a quantitative means to answer the question: *given life is present on a planet, can we detect it?* The detectability criterion is distinct from habitability: a world might be habitable, but could host life that is not detectable. The example of $H_2S$ above provides one such example, as do cryptic or marginal biospheres. Considering detectable biosignatures, by definition, should be much more likely to be observed being produced by living processes than non-living processes, one could consider Eq. (5) for D > 0 as a threshold for the quantitative definition of a detectable biosignature. More detectable biosignatures have higher values of *D*. It should be clear that a given observational signal may be a detectable biosignature in one environment and not another, depending on the value of *P(data|abiotic)*. This is related to the point made earlier that a given signal may be a biosignature, but not be evidence for life – this occurs if $D \le 0$.

In our framework*, "data"* can refer to different kinds of observations: the statistics from planet surveys, the context of a particular planetary system, or the observation of the planet itself. Here we will not be asking about the probability of the observation of a planet relative to instrument capabilities and distributions in the galaxy, as we leave this for the review by Fujii *et al.* (2017, this issue) and we are in any case interested in quantifying the likelihood of life on already identified targets. Instead, we focus on "data" with regard to direct observations of a planetary system and a planet that could host life. More often, a suite of these kinds of data will be utilized, where planetary statistics, as well as understanding from life on Earth, could provide theoretical support to interpret the direct observation of a planet, by contributing to the relevant conditional probabilities. For example, detection of a gas in an atmosphere requires a process-based model of that atmosphere to determine the contexts in which a certain mixture may be geochemically plausible and thus whether it is a signal of abiotic processes. The measured variables could be the NIR absorbance features ($NIR$), the planet's mass ($M$), density ($\rho$), orbital parameters ($o$) (for transiting planets), and the expected planetary elemental composition ($c$), which may be based on the star's composition. Catling *et al.* noted that some contextual parameters will depend on the presence of life and some will not (e.g., in general (excluding significantly advanced technological civilizations) we do not expect biology to significantly contribute to a planets mass), leading to differences in their treatment in a Bayesian framework. Of those listed here, only NIR absorbance features ($NIR$) is in general expected to depend on the presence of life. As such, $data = f(NIR)$. The remaining observables should therefore be considered as the context of the observation, and the likelihoods and priors must be conditioned on these. Thus, for example, $P(data|life) = f(M, \rho, o, c)$ and $P(data|abiotic) = g(M, \rho, o, c)$ are both functions of the planetary observables (these functions could also include stellar observables as well). Obviously it is a long way to go from the values of a planet's mass, density, *etc.* to predicting the observational signatures of life on its surface: hence the need for new cross-disciplinary collaboration.

To bridge observations to biosignatures, the surface chemistry, atmospheric mass, temperature profile, outgassing rate and photochemistry of a planet must all be modeled, along with any putative biological processes that could be occurring on its surface. The modeled atmospheres can then be compared to observed spectral features, and the plausibility evaluated of the biogenicity of the observations. If no plausible abiotic model can reproduce the atmospheric context for the gas at the same level of detection, but a model including life processes can, then we could conclude that the gas is biogenic. In such cases we should expect D >> 1.

A Bayesian approach requires good models for exoplanet properties *in the absence of life* to tightly constrain *P(data|abiotic)*. In many ways, this seems like it should be easier than building models of inhabited planets: removing the biosphere could significantly simplify models. But, we do not know what Earth would be like without life. To model Earth without life requires extrapolation from uninhabited environments on Earth, from worlds considered uninhabited, or from the identification and separation of biosphere processes from geological ones. Most of the input parameters to such models are not known. This includes the geochemical context of the atmosphere, and in the case of surface coloration features, the surface geology of the planet. The unpierceable 'flatness' of the VIS-NIR (0.6-2.5 µm) transmission spectrum of GJ1412b (Kreidberg *et al.*, 2014) ) illustrates that *not* seeing a spectral feature does not necessarily mean that a gas is not there. In fact, features unobservable with current technology or at wavelengths accessible to a specific mission may become observable given more sensitive measurements or through instruments capable of measuring different wavelengths  (e.g., for the case of GJ1214b; see Charnay *et al.*, 2015). In example, volatile molecule chemistry *outside* the major constituents

of Earth's atmosphere *in* Earth's atmosphere is not even known. These are observables, which could in the future be measured. If no plausible abiotic model can reproduce observed spectral features at a given detection level, but no models including life processes can reproduce observed spectral features, then no conclusions can be drawn about whether or not the spectral features are biogenic in origin. Just because a signal cannot be explained abiotically does not mean it is biogenic.

The most challenging parameter to constrain is *P(life)*. In the absence of a theory for life's origins we do not have a means to calculate this probability *ab initio*. There may be biospheres that are undetectable because the signal-to-noise is too low or because they do not produce a measureable signal, This is a problem of detectability (e.g., $D \leq 1$), which is distinct from the problem of estimating the probability of the *prior* occurrence of life. Attempts have been made to estimate *P(life)* within a Bayesian framework by Carter and McCrea (Carter and McCrea, 1983) and more recently by Spiegel and Turner (Spiegel and Turner, 2012). Both concluded that *P(life)* could be close to 1 or zero based on our current state of knowledge (one inhabited planet with a single origin for life) and that evidence for a second sample of life is necessary to distinguish the likelihood that life is common from the likelihood it is rare. This is of course the goal of the exoplanet life-detection community. The question is, *how can we develop the most effective strategies for searching for life, faced with the challenge that we have only trivial bounds on its prior occurrence?* One strategy is to focus on detectability, as noted above, since we can at least identify targets where we are most likely to detect life should it exist on a planetary surface.

*P(life)* will, in practice, be decomposed into probabilities for different living processes. For example the probabilities for oxygenic photosynthesis and sulfate reduction will, in general, be different. We do not know the frequency of planets with oxygen-containing atmospheres, although we can model this for abiotically produced $O_2$, and in the next 20 - 30 years we will start to have measured frequencies. There are many stages of evolution in the history of life on Earth (Bains and Schulze-Makuch, 2016; Maynard Smith and Szathmary, 1995), some depend strongly on prior history, and others have occurred independently within the branching history of life (e.g., multicellularity has evolved independently many times), although as far as we know all life on Earth shares common origins. Assumptions about what biological processes are happening within a given planetary context must be made with care. These should be informed by knowledge of potential evolutionary pathways in a particular planetary environment, as well as how many distinct environments a planet could potentially have on its surface. Even on Earth there is debate about the stages of evolution in the history of life, which may potentially confound our analysis when extrapolating to other worlds. This necessitates deeper connection between the exoplanet and evolutionary biology communities.

In what follows, we treat each relevant term in the Bayesian framework in turn. *P(data|abiotic)*, *P(data|life)*, and *P(life)*. *P(data|abiotic)* and *P(data|life)* are more readily constrained, we therefore first assess what is known and future directions for calculating these likelihoods, before moving to the harder problem of constraining *P(life)*. Toward the end, we provide an illustrative example of the Bayesian Framework and potential directions for informing search strategies.

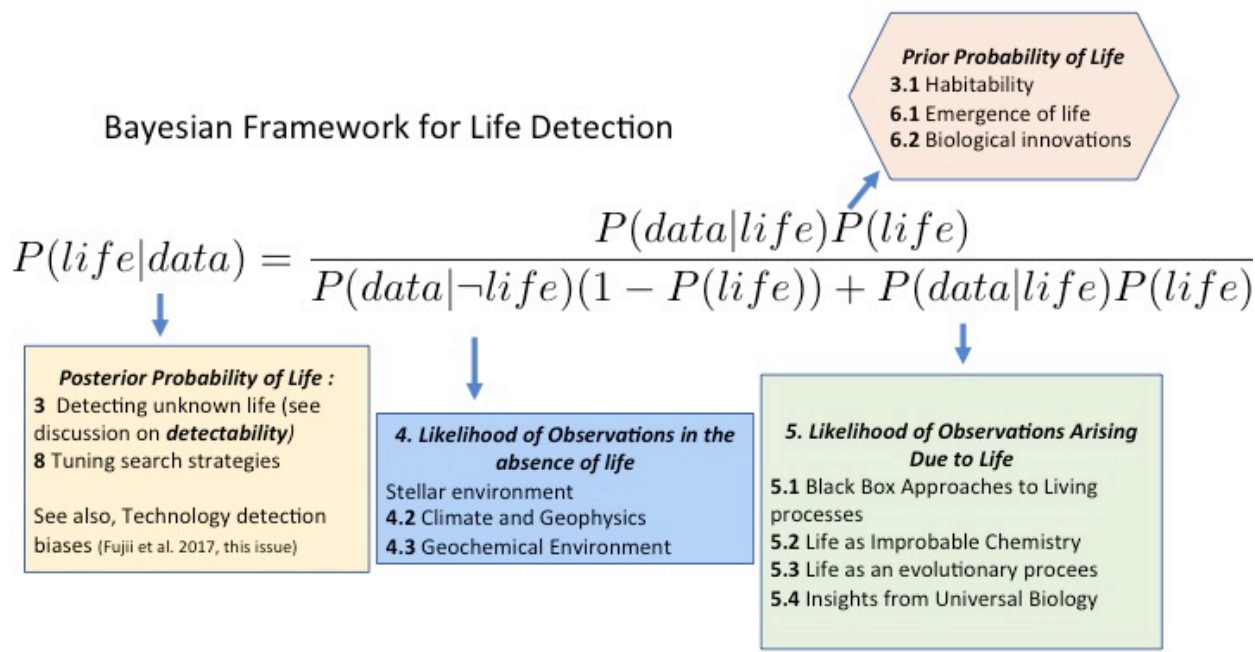

**Figure 1:** Conceptual diagram of a Bayesian framework for detection of exoplanet biosignatures, with section guides to this paper.

## 3.1 Habitability in the Bayesian Framework for Biosignatures

One of the most important metrics for guiding the search for life discussed within the exoplanet biosignature community is the concept of ***habitability***, where a habitable world is one where we expect Earth-life to be compatible. It is outside of the scope of this paper to provide a detailed discussion of habitability (see Schwieterman *et al*., Meadows *et al*., and Catling *et al*., 2017 this issue), or any ambiguities associated with definitions of habitability in relation to either life-as-we-know-it or life-as-we-don't-know-it. However, it is important to acknowledge the relationship between standard definitions of "habitability" and its relationship to terms in the Bayesian Framework.

The most commonly referenced definition of the ''habitable zone'', is the radiative habitable zone, defined to be that region around a star in which an Earth-like planet with an $N_2$-$CO_2$-$H_2O$ dominated atmosphere can have a surface temperature that could support liquid water (Kasting *et al*., 1993; Kopparapu *et al*., 2014). The concept of habitability implicitly makes assumptions about both *P(life)* and *P(data|life),* such that *P(life)$_{Earth-like}$* and *P(data| life)$_{Earth-like}$* > *0* within the 'habitable zone', where *P(life)$_{Earth-like}$* and *P(data| life)$_{Earth-like}$* > *0* are the prior probability for Earth-like life (by whatever definition) and the likelihood of observing the data given Earth-like life, respectively. The former is concerned with assumptions about the origins of life and its evolutionary innovations (discussed in Sections 6.1 and 7.2, respectively), the latter is concerned with life's ability to evolve and thrive in habitable environments (discussed in Section 5.3). Depending on the expectation of how habitability maps to the habitable zone, different priors can be constructed for *P(life)* as a function of radius from a star (and likewise for *P(data|life)*). If one assumes inhabited worlds to be limited to a habitable zone, then the assumption is *P(life) = 0* outside of the of the habitable zone (Figure 2A). If one assumes inhabited worlds are possible outside of the habitable zone, but much more likely inside the habitable zone, then *P(life) > 0* but small, outside the habitable zone, and *P(life) >> 0* inside the habitable zone (Figure 2B) (and could be such that *P(life)$_{Earth-like}$* > *0* in the habitable zone and zero outside). If one assumes the habitable zone is unrelated to the distribution of inhabited worlds, and life is equally likely at any radius from the host star, then *P(life) = constant* everywhere (Figure 2C). These are only a fraction of all possible prior scenarios (there are as many as there are hypotheses about the prior probability of life), and are given with the intent to help clarify how assumptions about habitability could translate to the quantitative formulation of biosignature assessment within a Bayesian framework. For example, assuming an inner radius around a host-star where conditions are too hot or radiating to allow habitable planets, implies *P(life) = 0* within that boundary. Currently, the value of *P(life)* in the habitable zone, or outside it, is not well-understood. *P(data|life)* is much better constrained, especially for oxygenic photosynthetic life (see Schwieterman *et al*., Meadows *et al*., and Catling *et al*., 2017 this issue for discussion of biosignature observables in the habitable zone). We discuss how to advance our understanding of *P(data|life)* to other scenarios for alien biospheres in Section 5, and *P(life)* in Section 6 below.

In this paper, we focus on the *detectability* of life, quantified in terms of likelihoods for biotic and abiotic signals, rather than habitability since the latter is discussed so extensively elsewhere (see Schwieterman *et al*., Meadows *et al*., and Catling *et al*., 2017 this issue). Importantly, we do not necessarily need to know what makes a planet habitable to identify planets with detectable biosignatures (although habitability can of course provide guidelines for detectability). Detectability is distinct from habitability: a world might be habitable, but could host life that is not detectable. Alternatively, a world may be "uninhabitable" (based on our limited understanding of planetary habitability, or the definition of habitable used, *e.g.,* lacking presence of liquid water on its surface) yet could host life that is detectable (for example, utilizing a different solvent than liquid water). This distinction between detectability and habitability allows us to, in this paper, expand the concepts of *P(life)* and *P(data|life) implicitly* underlying discussions of habitability and make them *explicit and quantitative*. By focusing on detectability, we hope the framework laid out in this paper will be useful for guiding the future directions of biosignature science, and will readily accommodate changes to the community's understanding of planetary habitability.

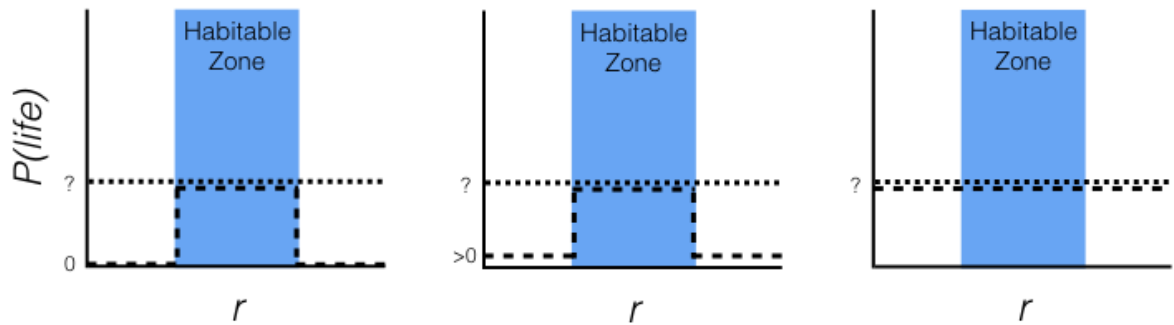

**Figure 2:** Examples of the relationship between the canonical definitions of the habitable zone and our assumptions about P(life), based on different priors. (**A**) A prior where habitability is limited to the habitable zone. (**B**) A prior where habitability is not limited to the habitable zone, but where P(life)$_{inside\_HZ}$ >> P(life)$_{outside\_HZ}$. (**C**) A flat prior where habitability is equally likely at any distance from the host star (not dependent on the habitable zone). By definition, the concept of a habitable zone implies that we expect P(life) > 0 (but of unknown value) for worlds within the habitable zone. See section 6.3 below for further discussion on P(life) and the habitable zone. This paper focuses on detectability as opposed to habitability. These examples are given to clarify how habitability might integrate into the Bayesian framework we outline above, but we do not go into further detail on this topic in this paper.

## 4   P(data|abiotic)

To reliably distinguish worlds with life from those without it, we must improve our understanding of worlds without life and their observational signatures. That is, we must constrain *P(data|abiotic)*. This is being pursued through modeling (Krissansen-Totten *et al.*, 2016; Domagal-Goldman *et al.* 2014; Harman *et al.* 2015; Luger & Barnes 2015; Schwieterman *et al.* 2016), but is a much more difficult problem observationally. To guarantee the absence of life, it would not be sufficient, for example, to make observations of planets outside of the habitable zone alone: those worlds may well be inhabited (see discussion in Section 3.1). Our assumptions regarding what worlds are likely to be uninhabited are most certainly incomplete, -- different forms of life may thrive in environments not compatible with our current concepts of habitability, e.g. subsurface life may somehow have an unexpected connection to the atmosphere. Furthermore, planets within the habitable zone that have no obvious "smoking gun" biosignature may nevertheless be inhabited, as exemplified by the early Earth which possessed a photosynthetically active biosphere, where net production and consumption fluxes balanced rendering atmospheric biosignatures challenging to detect (Reinhard *et al.*, 2017).

These examples make clear more work must be done to improve models to identify observational signatures of planets without life if we are to understand planets with life. This can be done through a combination of detailed understanding of abiotic processes, as developed from theoretical models, and observational surveys that select with care likely uninhabited worlds for observation to constrain *P(data|abiotic)*. *By better constraining the observables of strictly abiotic planets, it will become easier to disentangle true-positive biosignatures from false-positive biosignatures and to understand cases where life might be present, but not detectable.* Here, we focus on what is known and what needs to be known to determine *P(data|abiotic),* including constraining external planetary system parameters and internal planet characteristics in the absence of life. Each context considered – stellar environment, climate, and geochemistry – also impacts *P(data|life)* and *P(life)* as the likelihood and prior probability of life cannot be disentangled from its planetary context; we therefore also discuss these terms where appropriate.

## 4.1   Stellar Environment

Stars both influence planetary processes, and affect our ability to detect planetary properties, including any potential biosignatures. Catling *et al.* (2017, this issue) thoroughly summarize basic features of a parent star that influence or serve as indicators of a planet's atmosphere and potential development of life, including stellar age, effective temperature, composition (metallicity), spectral irradiance to the planet including flaring and particle flux, and whether it is part of a multiple-star and multiple-planet system. If we are to study the statistical probabilities for the emergence and likelihoods of life on different worlds, assessing the probability distributions of each of these stellar quantities throughout our galaxy is a key component, as each will affect the planet, influencing *P(data|abiotic)* and *P(data|life)*, and its potential to be inhabited, influencing *P(life)*.

Stellar surveys to characterize properties of stars of different masses, and hence temperatures, continue to add to our understanding of the potential impacts of stellar temperature on the search for life. The host star's temperature defines the radiative habitable zone, where we expect *P(life) > 0* (at least for life like Earth's, see Section 3.1). Astronomers are able to measure a star's temperature typically to better than 2-5% providing an accurate measure of the stellar irradiation, at least for wavelengths dominated by the star's Planck (black body) function. A star's temperature is closely tied to its mass, and we have strong constraints on the mass distribution in the stars in our galaxy (Reid *et al*., 2002; Bochanski *et al*., 2010; Bovy, 2017). The relative abundance of spectral types is much greater for cooler, long-lived stars, for which the habitable zone is closer to the star (0.1-0.4 AU). This makes for a higher probability of observing transiting planets in the habitable zone of cooler stars. The spectral energy distribution of a star's radiation will have different impacts on a planet's climate, due to the spectral properties of its atmospheric gas photochemistry and surface albedo, affecting potentially all three of terms of the Bayesian framework: *P(data|abiotic)*, *P(data|life)* and *P(life)*.

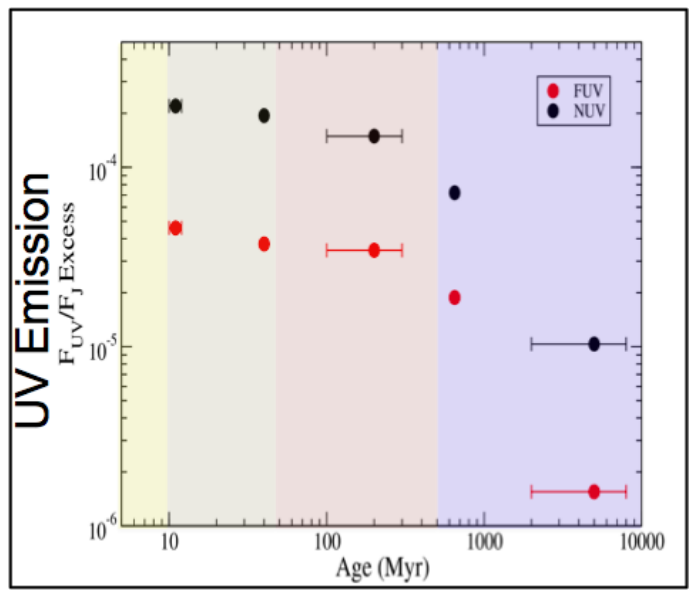

**Figure 3:** Median X-ray, far-UV (FUV), and near-UV (NUV) excess fractional fluxes , including upper limits, as a function of stellar age for early M stars. The radiation environment changes in time and is more intense for young stars, potentially impacting the probability of life emerging P(life). Figure adopted from Shkolnik and Barman (2014).

The lifetime exposure of planets to damaging stellar UV radiation is a key environmental factor for calculating the likelihoods and priors in the Bayesian framework. Increased stellar activity, through UV emission and associated particle flux can have dramatic effects on a planet's atmosphere (Segura *et al*., 2010; Luger and Barnes 2015). Studies have examined effects on the destruction and generation of secondary products of biogenic gases (Domagal-Goldman *et al.,* 2011; Segura *et al.,* 2005; Segura *et al.,* 2003; Hu *et al.*, 2012, 2013; Hu and Seager, 2014). While predicting atmospheric chemistry and biosignature gases through coupled radiative-convective/photochemical models is a mature method for Earth, atmospheric evolution of planets and the subsequent time-dependence around perpetually UV-active stars is not understood.

Unlike G-type stars like the Sun, M dwarfs are known to be active, with high emission levels and frequent flares when they are young, and this activity reduces as they age (see Figure 3; Shkolnik and Barman 2014). The large variability in M star UV outputs compared to Sun-like stars throughout their lives is just now being quantified and shows an increased level of activity towards cooler stars (Miles and Shkolnik 2017). The effects of sustained high levels of stellar activity on planetary atmospheres have not yet been studied, in part due to our lack of knowledge of UV flare rates and energies across stellar ages. However, efforts to resolve this are underway (Shkolnik et al. in preparation).

Explorations of the parent star role in planetary processes are recently expanding from 1D models to 3D General Circulation Model (GCM) based techniques. Rigorously quantifying atmospheric and water vapor loss can be informed through 1D models, but is dependent on magnetospheric shielding, which requires improved constraints through measurements and additional modeling. Interactive chemistry in GCMs for exoplanet studies is still in its early stages; few GCMs have the radiation capability to study atmospheric compositions that differ so substantially from modern Earth. In general, climatological GCMs can perform time slice equilibrium climate simulations given atmospheric composition (which may be provided by 1D models) or with photochemistry within Earth-like ranges. Conditions such as reducing atmospheres, absence of oxygen, condensation of greenhouse gases, and change in atmospheric mass at the edges of the habitable zone require further long-term model development. Thus $P(data|abiotic)$ is currently relatively unconstrained with respect to how stellar activity impacts atmospheric observables.

This is therefore is an important area for future research; M dwarfs have far-UV (FUV) to near-UV (NUV) flux ratios ~1000 times greater than the Sun (France *et al*., 2016; Miles and Shkolnik 2017), and represent 75% of stars. Small planets in the habitable zones of M dwarfs are common, and could have abiotic $O_2$ and $O_3$ levels 2–3 orders of magnitude greater than for a planet around a Sun-like star, due to hydrogen escape from stellar activity or photolysis of $CO_2$ (Domagal-Goldman *et al.,* 2014; Harman *et al.,* 2015; Luger and Barnes 2015). This is an example of a *false-positive* biosignature of oxygenic photosynthesis (Tian *et al*., 2014; Domagal-Goldman *et al*., 2014; Harman *et al*., 2015; see also Meadows *et al*., 2017, this issue). False positives suggest the presence of life, but occur where *P(data|abiotic)* is comparable to *P(data|life)* confounding interpretation of biogenecity.

Conversely, M stars may also become quiescent as they age such that they emit very little UV. The lack of UV to generate ·OH radicals can increase the detectability of biologically generated gases that would otherwise be removed by OH (Segura *et al*., 2005), increasing *P(data|life)*. It is therefore critical to determine the lifetime exposure of such planets to stellar UV radiation, from quiescent and flare emission levels, and explore the limitations on our ability to predict the resultant atmospheric properties.

In terms of detectability, we should expect that for most observables we might associate as biosignatures, *D>1* in some environments, but not others. For example, $O_2$ can accumulate to high levels on lifeless planets due to runaway water loss around pre-main sequence M stars (Luger and Barnes, 2015), as discussed above. The observation of collisionally induced absorption of $O_4$ (e.g., Misra *et al.*, 2014) would allow one to calibrate *P(data|abiotic)* for this

process. For example, one might set *P(data|abiotic)* close to 1, where the abundance of $O_2$ is the data, due to knowledge of these abiotic processes . In contrast, a planet in the habitable zone of a G-type star with properties similar to modern-day Earth (liquid water, ~1 bar of total atmospheric pressure, percent-levels of $O_2$, and relatively low levels of $CO_2$ and CO) strongly suggests a photosynthetic origin for atmospheric oxygen (Meadows, 2017 this issue). Again, evaluating *P(data|abiotic)* for the presence of $O_2$ requires contextual information about stellar environment, background atmospheric characteristics, and co-occurring atmospheric species, but in this case would yield *P(data|life) >> P(data|abiotic)*.

As we increase our knowledge of how planetary systems are influenced by stellar properties, through modeling and observations, there is a rich set of relevant phenomena to explore. Photochemistry interacting with radiation from different stellar types can inform our understanding of atmospheric chemical disequilibrium and detectable primary and secondary biogenic species, and research on the effect of the parent star's UV flares on prebiotic chemistry for the origins of life will be useful for constraining *P(life)* (Airapetian *et al.,* 2016).

## 4.2   Climate and Geophysics

The distribution of climate types and their variation in time results from star-planet orbital dynamics, and interaction between landmass and ocean configuration with circulation patterns. To address these nuances, in recent years 3D general circulation modeling (GCM) of rocky planet climates has emerged as a viable means to characterize circulation patterns on a planet and its potential to host detectable life (Leconte *et al.,* 2013; Way *et al.,* 2016). While 1D models remain extremely useful (Schwieterman *et al*., 2017, this issue), GCMs offer a tool to explore the variation in climate over a planet. Their strengths are that they offer self-consistent, spatially and temporally varying treatment of moist convection, clouds, atmospheric/ocean transports, and surface ice. They can be used to investigate the effects of obliquity (Abe *et al.,* 2005; Williams and Holloway 1982), eccentricity (Williams and Pollard 2002), and rotation rate, including tidal locking (Del Genio and Zhou 1996; Del Genio *et al.,* 1993; Edson *et al.,* 2011; Heng *et al.,* 2011; Joshi 2003; Joshi *et al.,* 1997; Merlis and Schneider, 2010; Pierrehumbert, 2011; Wordsworth *et al.,* 2011; Yang *et al.* 2014), providing a direct way to model the impact of exoplanet observables on climate, necessary to constrain the values of *P(data|abiotic)* (and also *P(data|life))*. Where 1D models are subject to extreme responses, the circulation patterns in GCMs generally have moderating effects (Shields *et al.,* 2013), broadening the expected range where $P(life)_{Earth-like} > 0$ compared to that predicted by average conditions alone (assuming $P(life)_{Earth-like} = 0$ outside of the canonical habitable zone). GCMs can be used to broaden concepts of super-habitability, defined whereby *P(life)* for a super-habitable planet is relatively larger than *P(life)* for Earth-like worlds, and habitability of less Earth-like planets (expanding the potential for life to worlds where *P(life)* would otherwise be assumed to be close to 0).

The role of ocean/continental configuration in influencing the distribution of planetary surface conditions has yet to be explored, with existing studies limited largely to either Earth's continents, or all land or aqua planet configurations. Some studies have experimented with having a planet with one hemisphere covered by land and the other ocean (Joshi 2003), continents at high or low latitudes at different obliquities (Williams and Pollard 2003), an equatorial super-continent, an aqua planet and planets with configurations similar to modern Earth continents (Charnay *et al.,* 2013). Life feeds back to a planet's climate by altering the composition of greenhouse gases in the atmosphere and changing surface albedo and water vapor conductance, which may reinforce or enhance the detectability of life. The potential of GCMs to characterize the extent and temporal variability of surface conditions remains to be explored. Future directions should add more realistic physics for alternative planetary contexts than Earth and focus on generating large statistics for the likelihoods of a given set of observations for both living and non-living worlds.

GCMs also offer a means to distinguish clouds from hazes (a potential biosignature) and to map climate zones over the planet's surface to surmise potential productivity, providing models to predict *P(data|life)*. For example, differential insolation on rocky planets can drive up-down circulations that cause large spatial differences in cloud cover and altitude, showing what windows through the atmosphere may be available to observe biosignatures for different stellar types and planetary rotation rates. Other questions to explore include whether a haze is universally a feature of homogeneous planets, or, in cases where atmospheric water vapor is detected, whether surface liquid water could be inferred through modeling, informing *P(data|abiotic)*. Given the large parameter space, a perturbed parameter ensemble approach is often used with Earth climate modeling and could be used to establish a library of a large number of GCM simulations covering a wide range of conditions. From this data set, the probability that observed properties arise from specific features such as clouds or hazes can be inferred, generating the large statistics necessary for getting tight bounds on both *P(data|abiotic)* and *P(data|life)*. Additionally, conditions conducive to observing biosignatures could be identified for target selection on future missions, or the large statistics may reveal patterns to classify planetary climates (Forget and Leconte, 2014).

Surface albedo plays a principal role in the surface energy balance of a planet, but exoplanet-observing missions in the near future will not be able to measure this directly. GCM studies typically prescribe the land surface albedo of a hypothetical planet to 0.1-0.3 (Abe *et al*., 2005; Abe *et al*., 2011; Wordsworth *et al*., 2011; Yang *et al*., 2014). However, mineral shortwave albedos can range from black volcanic rocks to white salts. A small change in albedo can significantly change climate. For example, for an instellation $S$ (W/m$^2$) and albedo $a$, the stellar energy intercepted per surface area of a planet is E = S(1-$a$)/4. Therefore, a change in $a$ of, say, d$a$=0.01 with $S$=1361 W/m$^2$ (the estimated solar constant; Kopp and Lean, 2011) gives an energy balance change of 3.4 W/m$^2$. There is currently lack of a theory for planetary evolution that would allow prediction of a planet's surface albedo or distribution of albedos, which is a necessary parameter for *P(data|abiotic)*. A community effort is needed to develop such a theory, which would depend on element abundances, processed by mantle melting, crystallization, the presence of water, and other system factors.

Other parameters difficult to constrain from observational data as well as theory include: atmospheric pressure, atmospheric mass, land/ocean ratio, land topography, and ocean depth. With near-term missions, it may be possible to measure obliquity, eccentricity, and rotation rate through photometric temporal variability (see Fujii, *et al*., 2017, this issue). Theory may also constrain rotation and obliquity in some cases: planets sufficiently far from their star will have had little tidal evolution, and rotation and obliquity will be hard to constrain from physics alone (Rodriguez, et. 2012).

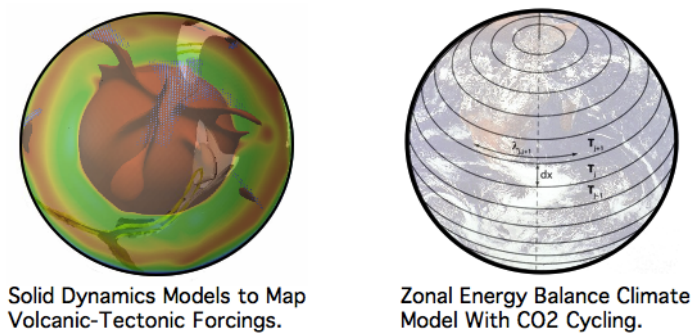

Solid Dynamics Models to Map
Volcanic-Tectonic Forcings.

Zonal Energy Balance Climate
Model With CO2 Cycling.

**Figure 4:** Modeling methodology used to explore the effects of variable volcanic-tectonic activity on planetary climate. Solid planet dynamic models of coupled mantle convection and surface tectonics (Lenardic *et al.*, 2016) are used to map out variations in volcanic and tectonic activity over time for a range of planetary parameter values (left image). Results from the solid dynamics models are then used to generate volcanic-tectonic forcing functions for zonal energy balance climate models (Pierrehumbert, 2010) that include volcanic degassing, topography generation, and $CO_2$ drawdown from the atmosphere due to surface weathering (right image).

### 4.2.1  Coupled Tectonic-Climate Models

In addition to surface properties, determining the composition and internal structure of exoplanets from orbital and transit data is moving toward statistical approaches (Roger and Seager, 2010; Dorn *et al.*, 2015). Composition sets the stage for quantifying life potential in terms of available biological building blocks and their likelihoods. How those building blocks are cycled over the geologic evolution of a planet to allow for conditions conducive to the development and evolution of life brings a temporal element, expanding the necessity of applying statistical approaches. GCMs are being used to investigate potential climatic states that may or may not be favorable for life. As GCMs perform time-slice equilibrium simulations, they are effectively instantaneous models when it comes to planetary evolution -- they do not track variable greenhouse forcing due to volcanic-tectonic activity on geologic time scales. Climate variability, on a Myr timescale or greater, is influenced by a greenhouse forcing that is modulated by a balance between the rates at which $CO_2$ is expelled from volcanoes and drawn down from the atmosphere via chemical weathering processes (Walker *et al*., 1981; Staudigel *et al*., 1989; Dessert *et al*., 2001; Coogan and Dosso, 2015). The global rate of $CO_2$ outgassing is governed by the character and pace of a planet's volcanic activity. Chemical weathering is mechanically paced by the rates at which new surfaces are created (Sleep and Zahnle, 2001; Whipple and Meade, 2004; Roe *et al*., 2008; Lee *et al*., 2013, 2015). The protracted clement climate of the Earth is, in part, a consequence of this long-term carbon cycle not having gone so far out of balance as to initiate a transition to a runaway greenhouse or a protracted hard snowball state. The degree to which this may be possible for planets beyond Earth, over a significant portion of their evolution, remains unanswered. Addressing that question has moved the community toward *coupled tectonic-climate models*, as shown in Figure 4.

Using coupled tectonic-climate models to address life potential will demand a statistical treatment given the number of parameters associated with coupled models and given the potential of planetary scale transitions over time. The capacity of the global climate of a planet to transition between multiple stable states has long been acknowledged (Budyko, 1969). Such climate transitions were initially investigated in terms of how orbital forcings could trigger them. However, volcanic-tectonic forcings can also trigger transitions in the climate state of a planet (Lenardic *et al*., 2016), and it has been argued the volcanic-tectonic state of a planet can itself also transition between multiple states (Sleep, 2000). The potential of bi-stable tectonic behavior (multiple tectonic states existing under similar parameter conditions) has now been demonstrated by several studies (Crowley and O'Connell, 2012; Weller *et al.*, 2015; Becovici and Ricard, 2016). Tectonic and climate transitions, over timescales of planetary evolution, bring historical thinking into the mix in a direct way for models that explore planetary conditions over time, permitting the possibility of constraining *P(data|abiotic)* and *P(data|life)* for different stages of planetary history and for different histories. This introduces the potential of variable paths for planetary evolution springing from initial conditions that can be very similar: acknowledging this in a modeling framework moves us away from a classical deterministic approach aimed at prediction. Instead, the objective is to map planetary potentialities in terms of their likelihoods, constrained within the bounds of physical and chemical laws. For example, a goal is to determine the likelihood a planet of a given size and composition (with uncertainties) orbiting a particular star in a particular orbital path, conditioned on a specified geologic time window, variable initial formation conditions and time variable climate forcings (orbital and/or volcanic-tectonic). Producing any kind of constraints on these planetary potentialities would yield a significant improvement in our ability to produce relative values of *P(data|abiotic)* and *P(data|life)* — and are crucial to maximizing detectability of biosignatures.

### 4.2.2 *Community GCM Projects for Generating Ensemble Statistics for P(data|abiotic) and P(data|life)*

Many of the planetary parameters to configure a GCM will not be measurable, or will be difficult to obtain given available observation technology, and require large computational resources. Furthermore, efficient sampling of the parameter space is necessary for climate sensitivity studies and generating statistical models. Constraining the parameter space theoretically is much needed and provides avenues for cross-disciplinary research. This may seem like a daunting task, but the initial steps are well within reach. GCM models, for example, can address long-term temporal evolution through ensemble simulations that capture time-slice equilibrium climates along some evolutionary path. For example, Way *et al.* (2016) performed model experiments run under different solar forcings associated with different points along the Sun's luminosity evolution. At the same time, models of volcanic-tectonic evolution have been progressively mapping potential volcanic-tectonic forcings that can be linked to simplified climate models (Lenardic *et al.*, 2016).

As the use of GCMs becomes more common to explore climates of exoplanets as well as of Solar System planets, model intercomparison studies will be necessary to gain confidence in their predictions. These complex models are subject to their own biases as a result of particular choices in numerical resolution and representation of physics. The Earth climate modeling community has coordinated projects for model intercomparisons that the exoplanet community may consider emulating. The Palaeoclimate Modelling Intercomparison Project (PMIP) (Saito *et al.* 2013; Joussame *et al.*, 1999; Pinot *et al.*, 1999) began in the 1990s, to compare studies of the Holocene. These studies also contribute to the Climate Model Intercomparison Project (CMIP), established in 1995 under the World Climate Research Program (WCRP) (Meehl, *et al.* 2000), which coordinates studies covering pre-industrial, current, and future climate scenarios. These experiments serve as important material for the Intergovernmental Panel on Climate Change (IPCC). The MIPs serve to define common climate scenarios, compile data sets for model inputs and evaluation, and agree on common model diagnostics to aid intercomparison (Eyring *et al.*, 2015). Modeling groups contribute ensembles of simulations that are archived for community analysis, providing insights into model biases, and strengths and weaknesses in scientific understanding of specific aspects of climate. The exoplanet community could utilize similar methods.

## 4.3  Geochemical Environment

As discussed above, the simplest approach for identifying a promising biosignature would be to search for a 'smoking gun' (which we discussed is unlikely to exist); something that on its own provides strong evidence for a biosphere (e.g., for which $P(data|life) >> P(data|abiotic)$). However, this type of signal is intrinsically vulnerable to 'false positives' as discussed in Section 4.1 (see also Meadows *et al*., 2017, this issue): contextual information about the geochemical environment is critical for accurately evaluating $P(data|abiotic)$. Another challenging problem is that of ***false negatives*** (Reinhard *et al*., 2017), or scenarios in which biological activity at the surface is overprinted by internal recycling and thus remains cryptic to characterization through atmospheric chemistry. Oxygen again provides an instructive example (see Meadows *et al*., 2017, this issue) - it may have taken hundreds of millions of years or more (Lyons *et al*., 2014) subsequent to the emergence of oxygenic photosynthesis on Earth before $O_2$ (or $O_3$) could be remotely detectable in Earth's atmosphere. The mechanisms underpinning the timing of this biogeochemical disconnect are still not entirely understood, but doubtless involve large-scale planetary processes unfolding on protracted timescales, such as hydrogen escape from the upper atmosphere (Catling *et al*., 2001), secular differentiation of Earth's upper crust (Lee *et al.*, 2016), and potentially a range of other factors. An important challenge moving forward will be to distinguish between the $P(data|life)$ values of false negatives and the $P(data|abiotic)$ of truly lifeless worlds for a range of potential biosignatures. This provides strong impetus for the development of robust models for the range of geochemical environments produced by sterile planets.

An alternative, or complementary, approach toward evaluating individual biosignature species is to search for chemical disequilibrium within a planetary atmosphere, or between an atmosphere and a planet's surface (Hitchcock and Lovelock, 1967). For example, it has become common wisdom that atmospheric chemical disequilibrium on a planet can be a strong indication of life (Lovelock 1965). However, free energy from stellar irradiance as well as from volcanic outgassing, tidal energy, and internal heat all lead to disequilibrium *even on a dead plane*t. Rigorous efforts to quantify disequilibria specifically associated with life are an active area of research. Different metrics that have been proposed, including kinetic arguments with regard to the power or fluxes required for maintaining disequilibrium (Gebauer *et al.* 2017; Seager *et al.* 2013; Simoncini *et al.* 2013), and topological measures of the directionality of chemical reaction networks in an atmosphere (Estrada, 2012). Krissansen-Totton *et al.* (2016) use a metric of thermodynamic disequilibrium for solar system planets, quantified as the difference between the Gibbs energy of observed atmospheric and (in the case of Earth) surface oceanic constituents and the Gibbs free energy of the same atmosphere and ocean if all its constituents were reacted to equilibrium, under prevailing conditions of temperature and pressure. This measure is able to show that Earth's atmospheric chemical disequilibrium is orders of magnitude greater than that of the other solar system planets, and is characterized less by the simultaneous presence of $O_2$ and $CH_4$ than by the disequilibrium between $N_2$, $O_2$, and a liquid $H_2O$ ocean. It is important to note that the diagnostic potential of this thermodynamic biosignature on Earth relies to some extent on being able to delineate both the presence and basic characteristics (e.g., ionic strength) of a surface ocean (Krissansen-Totton *et al.*, 2016), which provides another example of the type of broader contextual information required for evaluating both *P(data|abiotic)* and *P(data|life)*.

Interpreting atmospheric chemical disequilibrium as a biosignature depends very much on the geochemical and planetary system context. The disequilibrium may be tipped in different directions if the extant life primarily derives its energy from the available chemical disequilibrium or from an endergonic utilization of stellar energy for photosynthesis. An observed disequilibrium maintained by the star or photochemistry may also be interpreted as an **anti-biosignature**, indicative of available energy that is *not* being exploited by life. An antibiosignature is a signal that indicates the absence of life. The counter-argument to suggesting a given disequilibrium is an anti-biosignature is of course the evolutionary one that life on that world has not evolved mechanisms to exploit the relevant enery source; alternatively, the kinetics of consumption via microbial metabolism may be outpaced by abiotic production fluxes because it is limited by some other factor. Future exploration of disequilibrium metrics are needed to investigate other atmospheric compositions, unusual gases, surface (liquid bodies and rock) reactions, orbital temporal effects, planetary evolution pathways that affect outgassing and internal heat, alternative coupled ecosystem-planet interactions, kinetic metrics to deduce surface fluxes of biogenic and abiotic gases, and the uncertainties in determining species abundances, temperature, and pressure in future remote observations. Generating statistical data sets quantifying how different planetary parameters and living processes affect atmospheric disequilibria will place new constraints on $P(data|abiotic)$ and $P(data|life)$.

### 4.3.1   Anticipating the unexpected: Statistical approaches to characterizing atmospheres of non-Earth-like worlds

One approach that sidesteps the need to either define the biosignatures produced by life or the processes that produce them is to search for any signal that is *unexpected* from an abiological model of a planet. Recalling Eq. (5), we can maximize detectability by either maximizing the numerator, *P(data|life)*, or minimizing the denominator, *P(data|abiotic)*. Even if there is an extremely small probability that a signal is consistent with life, we can still identify it as a biosignature if we can demonstrate there is yet a smaller (perhaps zero) probability for the signal to be consistent with an abiotic origin. As highlighted above (and in Catling *et al.*, 2017, this issue) there are many challenges associated with modeling abiotic production of biosignatures on Earth-like worlds. The next frontier challenge to address is that most work so far has assumed we know what gases we are modeling, with a bias toward gases that are potential biosignatures for life on Earth. We must develop strategies to avoid this Earth-centric approach if we are to determine *P(data|abiotic)* and *P(data|life)* for the many worlds that do not fit the narrow box of Earth-like parameters.

Expanding beyond Earth-like worlds was the impetus behind Seager *et al.* (2016)'s *'All Small Molecules'* project. This project, explicitly aimed at volatiles that could be atmospheric signatures, seeks to determine all gases that could stably accumulate in any atmosphere. There are a very large number of such molecules, and so filters are necessary to reduce this to a manageable number. In their initial study, Seager *et al.* limited the data set to molecules with no more than six non-hydrogen atoms that were likely to be stable in the presence of liquid water. The size limit was imposed because the number of possible molecules goes up more than exponentially with the number of non-hydrogen atoms, and so this made the problem computationally tractable: seven- and eight-atom molecules could be added in future iterations. Water stability was required as any molecule made by life which diffuses to the atmosphere has to be stable to passage through the water in that life, and must be stable in the presence of oceans, rain etc. This is a constraint that could be relaxed if non-aqueous solvents were considered a realistic option for future searches for life.

The goal of the project is not the ~14,000 molecules in the initial database in itself; this is a starting point. The goal is twofold: to provide a database for future work on biosignatures and to provide a database of potential molecules to probe biochemical 'laws' proposed to govern life on other words. We discuss the first here, and the second in Section 5.4 on universal approaches to biosignatures below. To provide a database for future work on biosignatures, work currently planned includes estimating from thermodynamic and kinetic parameters those molecules that might be formed geologically, and hence would be weak as evidence for life, e.g., because *P(data|abiotic)* is nonzero and detectability is potentially low. For those molecules that are highly unlikely to be geologically formed on a planet, NIR signatures could be calculated to see if they are detectable. This requires thermodynamic and kinetic modeling of each molecule in a planetary context, calculation of NIR signatures, and integration of that with the atmospheric composition of the target planet. This is a substantial task in its own right, and rapid methods for estimating kinetic parameters, NIR spectral features etc. are a research goal for this program.

A major gap in atmospheric modeling pointed out particularly by the Seager *et al*. work is lack of measured kinetic data for reaction rates of the vast array of possible biosignature molecules with plausible atmospheric or surface components. Even thermodynamic data for nearly all the molecules in Seager *et al*.'s list (2016) have neither been measured nor accurately estimated. Moreover, solubilities in water are unknown, and atmospheric reaction chemistry and kinetics are unknown. As a consequence, modeling the atmospheric chemistry of these molecules will be an exercise in expert-informed guesswork. A major issue is not only that these measurements have not been done, but also that there is little community interest in carrying them out. Measuring the kinetics of gas reactions at different temperatures and pressures is exacting work, but is not rewarded by high-profile publications; at best, the data becomes one set of points in a large database and it is the database curators who get the citation. A topic for future research is therefore to find new technologies for making kinetic and thermodynamic measurements on gases, gas mixtures and solutions substantially faster and easier, so the collection of meaningful data sets becomes a single experiment in its own right rather than a decade-long program for an entire laboratory (an analogy is DNA sequencing, which can only be used to compile the databases mentioned above for evolutionary studies because sequencing a bacterial genome is now a high school project, and not, as it was 20 years ago, a feat worthy of a major publication).

## 5 P(data|life)

Moving from a product-based to processed-based search strategy will better bridge biosignature research with other active areas of research regarding universal features of life. Product-based biosignatures are practical: due to the limitations of current detection methods it is ultimately the chemical products of life that we will directly observe. To interpret data on chemical products and assess biogenicity, the likelihood of a given signature to be the product of life must be determined. A process-based approach is necessary to quantify these likelihoods. Therefore, to constrain *P(data|life),* we must understand the living processes generating a given observational signal. By better constraining observables based on the kinds of living processes present, it will be possible not only to detect life, but begin to infer its properties to achieve the longer-term goal of *characterizing* it on other worlds.

## 5.1  Black-Box Approaches to Living Processes

In the absence of knowledge about the processes of life on exoplanets, models typically assume biological sources based on production rates for biology on Earth. These are what we term *Input/Output* models of biosignatures. The steady-state concentration of any atmospheric gas is a function of its source and sink fluxes. For life detection, we are interested in inferring the existence of biological sources. Sinks can be studied using chemical models of atmospheric or surface chemistry and photochemistry. Nothing need be assumed about the internal workings of life on a planet – it is a *black box* that consumes some gases and emits other gases.

To enable expanding our search beyond looking for Earth-like life on Earth-like worlds, new approaches are necessary. One proposed framework from Seager *et al.* (2013) advocates classifying biosignatures based on the processes which produced them, with the idea to use this as a guide to whether those processes could be predicted to be different in a different environment. Their classification scheme still regards life processes each as a 'black box' of unknown mechanism. It is taken as a given that life requires free energy to operate, and mass to grow and replicate. Their classification therefore considers the potential inputs and outputs of the system that could provide energy and mass to that system. Potential biosignature waste products are considered as the output from processes that (1) capture chemical energy, (2) capture biomass, or (3) other processes. They conclude the first two can be constrained (if not predicted) by the chemistry of the planet, while the third cannot. Here we briefly summarize their classification scheme. While this scheme does not have consensus in the community, it serves as a useful jumping off point for further exploring process-based biosignature classification schemes.

### 5.1.1 Type classification of Seager et. al. (2013a)

*Energy capture (Type I).* Energy capture can be achieved through life's exploitation of chemical gradients in the environment, as well as through harvesting of light energy. Biogenic molecules signaling such energy capture include gases as waste products, as well as pigments that provide the mechanism for light energy harvesting (in some cases these could be classified as other below, due to their role in photoprotection). Examples of biogenic gases on Earth are $CH_4$ from methanogenesis, and $H_2S$ from sulfate reduction.

In principle, the Type I gas products can be predicted from knowledge of the chemical environment of life, thus providing a methodology for building statistical databases of expected products as a function of environment, needed for calculating *P(data|abiotic)*. For example, Seager and co-workers predicted that ammonia could be a detectable atmospheric biosignature on a terrestrial planet with a hydrogen-dominated atmosphere on the basis of the thermodynamics of the atmospheric and crustal chemistries (Bains and Seager, 2012; Hu *et al.,* 2013; Seager *et al.,* 2013a), combined with a hypothetical energy yielding metabolism in that environment, with $N_2 + 2H_2 \rightarrow 2NH_3$. One challenge for this approach is the diverse range of chemical environments on Earth, as illustrated by the production of a reduced waste product ($CH_4$) by life on Earth, which has a generally oxidized surface. On an 'averaged Earth', methane cannot be a Type I biosignature gas, but in reduced environments it can be produced as a byproduct of energy capture from methanogenesis or from biomass fermentation. Earth has many reducing niches today because it has biology producing oxygen. Electrons are conserved, and oxidation does not exist without reduction. Organic matter produced by oxygenic photosynthesis serves as the substrate for methane production during decay processes. Buried organic matter then creates anoxic subsurface environments conducive to the production of methane. Future work should take the diversity of coupled environmental sources and sinks into account. The values of both *P(data|life)* and *P(data|abiotic)* may not only be a function of the bulk composition of a planet, but also the number and variety of distinct environments on its surface (see also Scharf and Cronin (2016) for a discussion of the role of diverse environments in potentially increasing *P(life)*).

Seager *et al.* (2013b) have demonstrated that it is possible to extend the Type I concept to estimate not only whether a gas could be the result of exploitation of a redox disequilibrium on a planet, but also whether that source is a plausible source of a *detectable* biosignature gas. The pilot study of Seager *et al.* suggests that further research on life's need for energy would help to focus which Type I products are plausible as detectable biosignatures.

*Biomass capture (Type II).* The carbon on planet-sized bodies with thin (Earth-, Venus- or Titan-like) atmospheres is likely to be mostly oxidized ($CO_2$) or mostly reduced ($CH_4$), as these are thermodynamic minima for carbon in an oxidized or reduced environment, respectively. Life needs to convert this into carbon in intermediate redox states to build complex molecules; this is a chemical universal, deriving from the nature of chemical bonds to carbon (Bains and Seager, 2012). This requires the oxidation or reduction of an environmental material, respectively. The input is an environmental chemical and environmental carbon, the output is biomass and a material out of thermodynamic equilibrium with the abiotic environment. The possible inputs and hence outputs are, again, predictable in principle allowing the possibility of constructing probabilities for the products of biomass capture, informing *P(data|life)*.

The case of photosynthesis illustrates both the power and the limitation of considering a whole organism as a black box, which considers only looking at the net reaction rather than the individual components being reacted (for example, in a biochemical pathway). Considered as a whole, the net mass balance for an oxygenic photosynthetic organism can be expressed as input of $CO_2$, $H_2O$, and light and output biomass and $O_2$ as an oxidized waste product of the reductant $H_2O$:

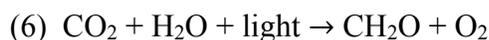

(6)  $CO_2 + H_2O + light \rightarrow CH_2O + O_2$

Here, $CH_2O$ is a simplified representation for a carbohydrate or sugar. In this context, oxygen is the principal output gas, and is Earth's most notable Type II biosignature (indeed, Earth's most notable biosignature of any sort). Similarly, other oxidized (non-gaseous) waste materials are generated by anoxygenic photosynthesis (see Kiang *et al*., 2007a; Schwieterman *et al*., 2017, this issue). The same logic can be applied to any life in any environment – indeed in principle the same logic could be applied to non-carbon-based life.  Using just the overall stoichiometry and thermodynamics of the net process of taking in environmental carbon and energy and outputting biomass, other oxidized (Haas, 2010) or reduced (Bains and Seager, 2012) products have been suggested for other worlds. Thus this 'organism-level black box' approach has power in providing a framework for suggesting overall inputs and outputs before understanding of internal mechanism is available for any biomass capture process.

*Other uses (Type III).* Life on Earth produces a wide range of volatiles for signaling, defense and other functions. Here a 'black box' approach is relatively powerless as an explanatory tool as these processes are highly contingent. There is no known way to predict what signaling or defense chemicals an organism will make, starting only from the overall physics and chemistry of its environment. The production of Type III gases is a result of the ecological or physiological demands on the organism, themselves the result of evolutionary contingencies and of relationships with other organisms: data that is not accessible for exoplanets. As a result, in principle we might consider any chemical to be a Type III biosignature, which was the motivation behind Seager *et al.*'s compilation of all possible small molecule, volatile biosignatures (Seager *et al.,* 2016).

*Products of modification of gases (Type IV).* Gases produced by life can be modified by the environment, providing a source of secondary signatures of life. Examples include ozone (the photolytic product of oxygen) and DMSO (the oxidation product of DMS). These could in principle be predicted if the environment and products of life are known, e.g., for Type I and II biosignature gases, but are will not be predictable for Type III.

## 5.1.2  Alternatives for Type classification

Seager *et al.*'s original classification (2013a) was introduced as a pragmatic approach to the specific task of extending our understanding of the input/output model of biosignature generation, permitting moving beyond basing models solely on terrestrial production rates. Suggestions at refinement do not necessarily converge on agreed classification systems, suggesting that there may be no exhaustive categorization method. The goal of devising such process-based classification schemes is to probe **why** life might evolve to produce a given chemical signature, as the Seager *et al.* classification was explicitly devised to do. A process-based classification particularly is needed for systems modeling to simulate the production of biosignatures as well as to explore possible novel biosignatures resulting from complex interactions along the entire pathway from metabolism through biosynthesis to post-processing in a given planetary biogeophysicochemical system. This approach is also useful for formalizing conditional probabilities in a Bayesian framework as it can inform the likelihoods of producing a particular gas conditioned on a given environmental context.

Other disciplinary perspectives pose alternatives to the Seager *et al.* approach, expanding our ability to calculate *P(data|abiotic)* and *P(data|life)* based on black-box methods. For example, while the Seager *et al*. Type classes focus on gaseous biogenic products and their secondary products in the environment, the classification could be further generalized to include surface biosignatures that also result from these Type processes, or properties of the chemical networks that generate them (see Section 5.4.1 below). Possible surface biosignatures include pigments, or even morphological features. Suggestions for generalizing and making the Type classifications more precise include the following:

*Type I, Energy capture*.  Light harvesting pigments can be included as a Type I biosignature molecules; although they are not the products of energy capture, they are the means to energy capture. These include pigments of oxygenic photosynthetic organisms, bacteriorhodopsin of Archaea, and other light absorbing molecules, as summarized in Schwieterman *et al*. (2017, this issue). Fluorescence as a result of excess energy release or waste product from light harvesting could also be considered a Type I biosignature. Moving from the level of individual molecules to the networks of their interactions, certain chemical networks may be better at energy capture than others, suggesting yet another metric for assessing the potential for life (see Section 5.4.1 below).

*Type II, biomass capture*. Biomass capture could be elaborated further.  The biomass itself can be a biosignature and produce waste products that also can be biosignatures. Biomass capture can be through autotrophy (reducing inorganic carbon, $CO_2$) but also through heterotrophy (incorporation of already reduced carbon).  Reduction of $CO_2$ into biomass does not necessarily produce waste products immediately, but only after that biomass itself is involved in other activities.  Incorporation of inorganic carbon into biomass can be highly complex, and the "black box" approach to metabolism can be insufficient for identifying biosignatures. If, for example, it was found that the use of reducing equivalents generated by photon capture to reduce $CO_2$ *necessarily* produced by-products or required other detectable properties of an organism, then these would be candidate Type II biosignatures (see Section 5.1.3 below for more details on the subtleties of Type classification). This requires an understanding of what aspects of photosynthesis are requirements of the steps in the chemical processes, and which are evolutionarily contingent, posing challenges for constraining *P(data|life)* (see Section 6.2 below).

Another candidate Type II biosignature is seasonal oscillations of atmospheric $CO_2$ content due to shifting balances of autotrophic carbon fixation versus respiration.. This is exemplified by the observed ~2% seasonal amplitude in $CO_2$ concentration in the northern hemisphere caused primarily by the growth and decay or senescence of land vegetation (Keeling, 1960; Keeling *et*

*al*., 1976, 1996; see Figure 12).

*Type III, "Other uses"*. The Type III classification is a catch-all to describe compounds whose production is not predictable on the basis of planetary chemistry, and as such includes biosignatures with diverse functions other than metabolic ones. These can include ecophysiological functions such as nutrient capture and heat tolerance (e.g. isoprene), ecological functions such as anti-bacterial and anti-microbial compounds (e.g. methyl bromide), and intra- and inter-organismal signaling (e.g., ethane). Schwieterman *et al*. (2015) summarize a variety of ecophysiological and ecological functions of numerous types of biological pigments, which include phototrophy, anti-oxidants, photoprotection (screening), thermal tolerance, nutrient acquisition, growth regulation, and ecological functions like antibiotics and signaling. Even non-chemical properties of organisms, such as bioluminescence, used for signaling, can be classed as 'Type III' (see Schwieterman *et al*. 2017, this issue). The Type III classification is therefore one that provides a "catch-all" for cases where the biosphere (considered as a black box) modeled only in terms of metabolic processes cannot be predicted to produce that signature. This is important in estimating the probabilities *P(data|life)*: for Type III biosignatures our uncertainties in biological origins mean that the signal-to-noise will not be sufficient for unambiguous detection and more context will be necessary to confirm biogenecity.

Type III could be extended in two ways. It could be subdivided into biosignatures likely to be produced for specific purposes. An example are retinal pigments that are unlikely to be used as visual signaling molecules in an exclusively microbial biosphere, although they may function in related activities such as phototaxis or ion transport. It could also be divided into functional classes in cases where modeling of a biosphere was sufficiently sophisticated to infer functional roles for particular class of signatures.

*Type IV,* Products of modification of gases were originally considered to be the products of environmental modification of gases, which in turn produce other gases. This could be generalized to products of environmental modification and degradation of biogenic molecules, including gases, liquid, and solid molecules to produce other gases, liquids or solids. For example, the terrestrial "Black Earth" Chernozem soils are the result of substantial modification of local geology by biology, and would not be found on an uninhabited world.

Some biosignatures may be of more than one Type. For example, marine algae produce dimethyl sulfide (DMS) as a byproduct of the breakdown of a complex biochemical dimethylsulfoniopropionate (DMSP). DMS can be classified as a Type IV product. However, it is also probably the principal energy source for the predatory zooplankton that feed on DMSP-containing algae, and so for them it produces a Type I biosignature (see Schwieterman *et al.* 2017, this issue, for more details). There is some debate what Type $O_2$ should be classified as, depending around what process(es) one draws one's black box. For example, Seager *et al.* (2013) classified $O_2$ as a Type II biosignature involved in carbon or biomass capture, for reasons outlined above: considering the organism as a single system, photosynthesis involves input of $CO_2$ and light, production of biomass and output of $O_2$. However, oxygen is produced as a result of oxygenic photosynthesis, which is achieved through several steps in series. In the "light reactions," photon energy is used to acquire electrons from water, whereas biomass capture through fixation of carbon from $CO_2$ occurs in a separate subsequent step in the Calvin-Benson-Bassham cycle, which is the same process used by anoxygenic phototrophs. These steps are detailed in Schwieterman *et al.* (2017, this issue). Photon energies in series drive successively more oxidized states of the oxygen evolving complex (OEC), a highly oxidizing metallocluster that, upon reaching a critical state, catalyzes oxidation of water, thus generating oxygen. Part of that photon energy is further used to excite the electron to a lower redox potential, the energy of which can then be used in redox reactions for storage of that energy. Photosynthesis self-generates its own chemical gradients, both to acquire electrons from water, as well as to support redox reactions for energy storage. In this more detailed analysis, oxygen is a byproduct of the step of energy capture and excitation of electrons; it is in effect the byproduct of capture of energy from an internally generated redox gradient, and so is a Type I product.

### 5.1.3  When is it appropriate to deconstruct a black-box?

The black box method can be used as a first approach to understanding biosignature production when the underlying biological mechanism(s) are unknown. However, the final example in the previous section highlights the potential ambiguities involved in such a classification scheme. Further challenges arise when the black box fails to work entirely, and more detailed resolution of the mechanisms 'hidden' in the black box may be necessary.

For example, Eq. 6 is a net mass balance equation for an oxygenic photosynthetic organism, in which similar terms on both sides of the equation have been canceled. It does not express the stoichiometry when other oxidized (non-gaseous) waste materials are generated by anoxygenic photosynthesis. Applying a black box to the inputs and outputs for anoxygenic photosynthetic organisms reveals some terms that cannot be canceled, which can then motivate further investigation to explain these extra terms. As just one example, drawing black boxes around many similar organisms and finding that some things about them are common while others fail to fit the same net reaction model would motivate dissection of the mechanism(s) of photosynthesis (dividing the black box into smaller black boxes, defined by our understanding of mechanisms). This is necessary, for example, to generalize for both oxygenic photosynthesis and the several types of anoxygenic photosynthesis,.

It turns out that photosynthesis involves processes that separate the activities of energy capture and biomass capture in sequence, so a more generalized black box is expressed in the equation:

(7)     $CO_2 + 2H_2A + light \rightarrow CO_2 + 4H^+ + 4e^- + 2A \rightarrow CH_2O + H_2O + 2A$

The intermediate reaction reveals that the reductant $H_2A$ must first be split to donate electrons, and the $CO_2$ is reduced subsequently. When the reductant is $H_2O$ then the equation yields:

(8)     $CO_2 + 2H_2O^w + light \rightarrow CH_2O + H_2O + O^w_2 \text{ (gas)}$

where the superscript "*w*" denotes that the molecules in the produced oxygen gas come from the water molecules and not from the $CO_2$. In the simpler black box equation, the $H_2O$ on the right-hand side canceled with one on the left, but in fact it is not the same, since the source of the oxygen atom is different. If the reductant is instead, for example $H_2S$, in anoxygenic photosynthesis, then the net reaction is:

(9)    $CO_2 + 2H_2S + light \rightarrow CH_2O + H_2O + 2S$ (solid)

This shows that specific waste products can be viewed as obeying a common set of processes. Both the $O_2$ and S are the results of oxidation of an input reductant from input for light energy to obtain electrons. In this context, they can be seen as the Type I product of energy extraction from an internal redox gradient. Even our last generalized equation is still a simplistic black box, depending on one's question. For example, as discussed in Schwieterman *et al.* (2017), this equation does not reveal how the light energy is partitioned, and in fact only a fraction of it is used in oxidizing the reductant. If addressing what wavelengths of light can be used in different types of photosynthesis, the black boxes must be dissected further. The reader is directed to Schwieterman *et al.* (2017), which provides more details and literature. Understanding more of the process by which photosynthesis occurs changes our perception of how oxygen is produced. ("Why" it is produced, the final cause in an Aristotelian sense, then depends on how one asks the question)

The example of photosynthesis exemplifies the value of an interdisciplinary discussion to address biosignatures, wherein the approaches of physicists, chemists, and biologists are together leveraged to identify the useful level of parsimony versus complexity. The power of the black box type of input/output model of life is that it can be implemented on the basis of environmental parameters alone. This is also its limitation, in that it says nothing about process or mechanism – it assumes that these are unknowable at interstellar distances. It remains an open question to what extent this last point is indeed true, and to which either universal 'laws of life' or more detailed understanding of the necessary chemistry of specific processes could unpack the black box (both discussed more extensively below).

## 5.2 Life as improbable chemistry

A different kind of "black box" approach focuses on the complexity of chemical products of life, rather than classification of how they are produced. One major observable that discriminates living things from inanimate matter is their ability to generate similar, complex or non-random architectures in large abundance, or to affect the background. Abiotic distributions of organics tend to be smooth, and dominated by low-molecular weight species, whereas in life, natural selection yields distributions that are more "spikey" as a result of selection of functional sets of molecules (Lovelock 1965, McKay, 2011). Life also reliably produces high molecular-weight biopolymers, whereas abiotic processes do not. Relating to the Bayesian framework, the idea of searching for low probability chemistry is the same as guiding our search for life by high detectability $D$: we should look for life where we expect no abiotic system could produce such a signal.

One potential biosignature could be the entropy of a distribution of molecules, distributions that are very unlikely to occur abiotically (*e.g.,* ones that require natural selection) are less probable. In this case, the biosignature is itself the probability of a molecule or a distribution of molecules occurring abiotically: if the probability is very low (low entropy) we can be confident the signal arises due to life. Caution must be taken in assigning biological origins to non-random processes, however. An example is the periodic distributions of masses of peptides displayed in living organisms, which have a mathematical rather than biological explanation: rather than being a product of natural selection, this pattern can be shown to arise purely as a result of the properties of finite ordered sums combining 20 natural numbers (corresponding to the 20 or so biological amino acids) (Hubler and Craciun, 2012).

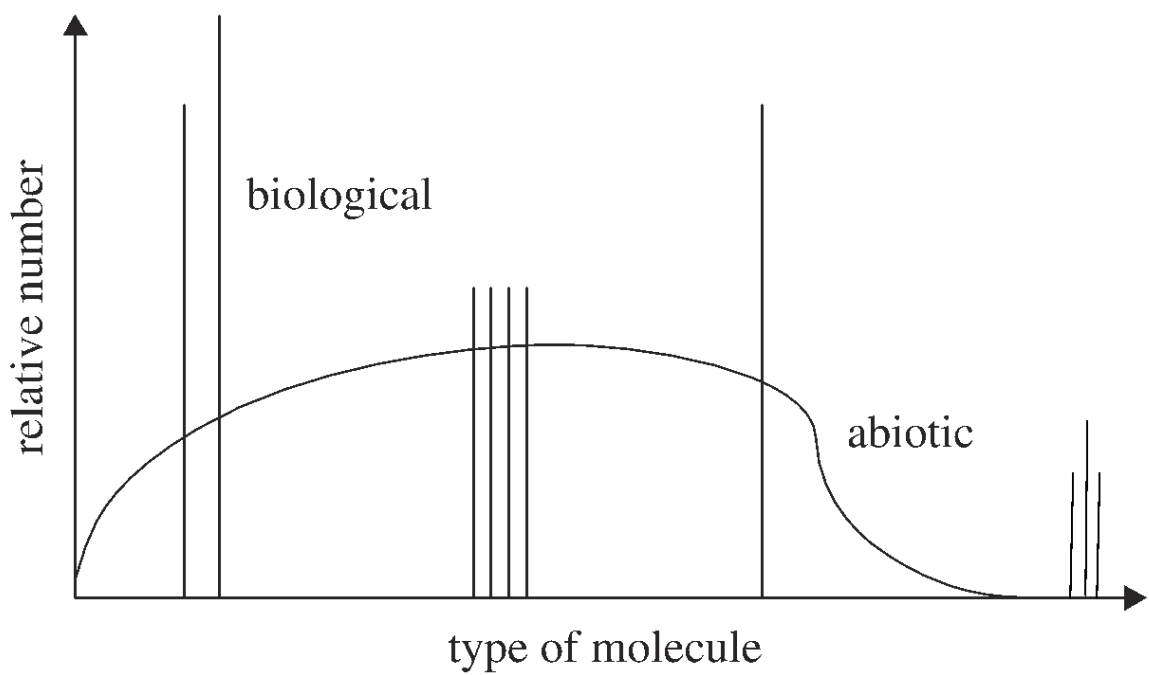

**Figure 5:** Schematic illustrating the difference between abiotic (smooth curve) and biological (spikes) distributions of organic molecules. Non-living systems tend to produce smooth, thermodynamic distributions, whereas in living processes only a subset of molecule species are selected (through natural selection) to form a functional set. Figure from McKay (McKay, 2011).

Based on the probabilistic framework, complex artefacts are themselves biosignatures, since they can potentially be discriminated from an abiotic background. For example technetium (Tc) is a rare element, not produced naturally, and has been proposed as a possible indicator of a technological civilization, since creating it requires knowledge of nuclear physics (Paprotny, 1977; Whitmire and Wright, 1980). Computers are also biosignatures for the same reasoning, it is unlikely that a laptop would form spontaneously without the long sequence of evolutionary steps necessary to evolve intelligence capable of constructing such technology. While highly improbable structures are most often discussed as biosignatures in SETI research (Sagan and Shklovskii, 1966; Drake, 1965) the concept of improbability is equally applicable to molecular signatures of life. The challenge is that we must threshold a minimum complexity above which we can be confident $P(data|life) > P(data|abiotic)$ for the molecule of interest.

It is not obvious how it might be possible to generalize an approach that aims to evaluate complex objects as possible biosignatures, particularly with respect to chemical signatures of life. In the context of exoplanet searches for life, achieving this on a planetary scale requires the probabilistic search for anomalies that themselves have in-built structure. Marshall *et al.* (2017) have recently developed one possible complexity measure, they call *Pathway Complexity*, which quantifies the complexity of a any given object as the shortest pathway for its assembly. The measure identifies the shortest pathway to assemble a given object by allowing the object to be dissected into a set of basic building units, and rebuilding the object using those units. Pathway Complexity bounds the likelihood of natural occurrence by modeling a naïve synthesis of the observation from populations of its basic parts, where at any time pairs of existing objects can join in a single step. An object of sufficient complexity causing an observable feature, if formed in the absence of life, would have its formation competing against a combinatorial explosion of all other possible features that are equally probable. Pathway Complexity can be seen as a way to rank the relative complexity of objects made up of the same building units on the basis of the pathway, exploiting their combinatorial nature. The motivation for the formulation of Pathway Complexity is to place a lower bound on the likelihood that a population of identical objects or observations could have formed or occurred abiotically, i.e., *without the influence of any biological system or biologically derived agent*. Thus, it is assumed that objects with high Pathway Complexity will only be observed if produced by life. Conceptually, the measure is similar to Bennett's logical depth, a measure of complexity based on the number of computational steps necessary to recreate a piece of information (Bennett, 1988). The key difference is that Pathway Complexity looks at the *intrinsic* routes for connecting objects based upon the resources within the system. Locally, Pathway Complexity could be used to rank molecules in order of complexity, enabling identifying a threshold above which the molecule must have been produced by a biological system.

For exoplanets, we are unlikely to remotely detect large macromolecules, and the Pathway complexity for remotely detectable small molecules is in general low. In principle, exceptions to low Pathway Complexity could include molecules such as dimethyl sulfide (DMS) and dimethyl disulfide (DMDS). These volatile gases are produced as indirect metabolic and decay products of both eukaryotic and prokaryotic organisms, require several independent enzymatically mediated steps to produce, and consequently have no known abiotic sources (see overview in Schwieterman *et al*., 2017, their section 4.2.5). In its current formulation, Pathway complexity cannot account for biogenecity of small molecules such as DMS and DMDS, as it is necessarily a combinatorial measure to be computable from only knowledge of the object. Since DMS and DMDS are below the threshold complexity set for origins from living processes (above which we can be confident life produced it), they would have low Pathway Complexity despite their complex biological synthesis pathways. Additionally, these specific example gases are unlikely to build up to detectable levels in planets orbiting stars other than inactive M dwarfs, but could be indirectly indicated by a $C_2H_6$ over abundance relative to that expected for the photochemical processing of $CH_4$ (Domagal-Goldman *et al.*, 2011).

Despite current limitations, the concepts driving Pathway Complexity are promising for exoplanet research, and there are potential ways forward for characterization of small molecules. Expanded versions of pathway complexity can account for the occurrence of many small molecule species occurring simultaneously, by analyzing the number of possible network pathways. This might provide fruitful new directions for assessing biological origins, or at least bound *P(data|life)*. More broadly, Pathway Complexity should be able to be used to connect spectroscopic signatures looking for patterns of improbable 'complex' behavior. An example could be exoplanets that have Complex 'LED-like' spectroscopic signatures that occur in abundance yet cannot be explained without technology. The key to expanding this will be using the Pathway Complexity to develop thresholds that are accessible by current technology or inspire the development of new technologies, experiments and approaches.

## 5.3   Life as an evolutionary process

The "black box" approaches of the previous sections are not concerned with the specific mechanisms mapping the planetary input to biological output, which is both their strength and primary limitation. The internal mechanisms of biological processes inside the 'black box' are driven by evolution, necessitating a better understanding of the universals of evolution to determine the universals in an input/output framework. In applying Earth life as the standard of reference, we have only one past and one present to guide inferences of how integration over microscopic effects produces specific macroscopic biosignatures. One challenge is disentangling contingent events (which take the form of temporal conditional probabilities) in the evolution of life on Earth from universal constraints that we might expect to apply to life anywhere. Using Earth's history of evolution, we are effectively substituting timing, frequency, and diversity for *likelihoods* applicable to the study of exoplanets.

The universality of the genetic code and central dogma indicate all known life on planet Earth originated from a last universal common ancestor (LUCA) (Koonin and Novzhilov 2009). However, components of LUCA, such as archaeal and bacterial cell membrane components and metabolic capabilities, may have arisen independently multiple times. Our understanding of early evolution is complicated by the common occurrence of horizontal gene transfer (Lombard *et al*., 2012; Woese, 2004, Mushegian 2008). For example, respiratory chain components needed for aerobic respiration may have been laterally transferred between bacteria and archaea, and the origins of these genes are unclear (Kennedy *et al*., 2001, Boucher *et al*., 2003). As a result, it is conceivable the last common ancestor of each extant gene may or may not have been present within the LUCA population. With respect to the origins of life on Earth, the existence of a LUCA implies our sample size is *N=1*, which does not provide enough data for statistical inference about the processes or likelihood of abiogenesis (see Section 6.1). However, the evolutionary process on Earth has driven innovations over many temporal and spatial scales, permitting the possibility of understanding more universal features of evolutionary processes by studying many events. Generalities could then be extrapolated to other chemistries. Thus, when taken in light of the diverse biogeochemical contexts for life on Earth, we might consider that we have *N=many* examples for calculating *P(data|life)* for evolutionary processes, rather than being restricted to *N=1,* as is the case for the origins of life (see discussion on *P(life)* below for cautions in assuming independence of evolutionary innovations for constructing likelihoods).

Extinction has been a hallmark feature of life on Earth. Therefore, in calculating *P(data|life)* we must not only consider the probability of an evolutionary innovation emerging, but also its persistence in time. It would be worth knowing whether a planet being observed has extant life or, if not, is otherwise suitable for the emergence of future life, or if observables indicate that life was once present in the past. Better understanding of evolutionary processes and how they couple to planetary scale signatures is necessary to make progress on these unknowns.

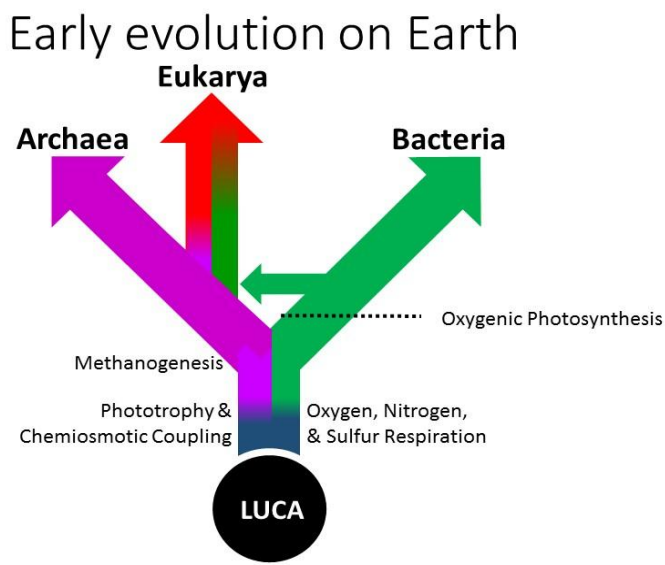

**Figure 6:** Life on Earth radiated from a last universal common ancestor with a single standard genetic code. With respect to core biochemical components of life, this common ancestry leaves a sample size N=1. However, the subsequent evolution of diverse metabolic capabilities and evolutionary lineages has resulted in diverse trajectories allowing the possible of mapping N=1 to N=many, considering the varying coupled environmental and biological states over geological timescales and the number of independent, convergent evolutionary innovations and transitions (only a few of which are shown for illustrative example). It is unknown how frequently these evolutionary events, including the origin of life, should be expected to occur on other worlds.

### 5.3.1  Life as a co-evolution with its planet: Earth as an example

In the years since Des Marais *et al*. (2002), the community has increasingly recognized the fundamental importance of understanding biosignatures as resulting from the coupled evolution of a planet with the life upon it. Life is a phenomenon that manifests and maintains itself at molecular and microscopic scales, and, through evolutionary processes tightly coupled to geochemical cycles, has led to macroscale changes in the Earth-system. Again, using atmospheric $O_2$ as the standard example, oxidation of water by photosynthesis occurs at the molecular scale, and the evolution of oxygen and the growth of organisms may be observed in the lab and field *in situ*, but the expression of $O_2$ as an exoplanet biosignature requires planetary-scale accumulation over geologic time scales. Due to the co-evolution of the biosphere, lithosphere, hydrosphere and atmosphere of Earth, life itself may be considered as a planetary process (Smith and Morowitz, 2016).

Planets are not static, but evolve in response to stellar context and planetary feedbacks: for living worlds, these feedbacks include those between a biosphere and geosphere. As evidenced by Earth's transitions, it is possible for precursors to biotic processes to emerge on an abiotic planet, mediating the transition to a living world, and for a living world to transition through different phases. Because Earth life has been exclusively microbial for the majority of Earth's history, it is possible that primitive unicellular life forms are the most common and longest lasting stage of life on a planet (Whitman *et al*., 1998). Microorganisms are thought to have evolved on Earth after catalytic and genetic macromolecules were compartmentalized into membrane envelopes (Lombard *et al*., 2012). The degree to which these early life forms may have metabolic activities like modern microbes, such as lithotrophic or photosynthetic processes, is currently unknown. It is still not well understood how feedback between the biosphere and the geosphere shaped the gases that would have been detectable in our own atmosphere during the period after life first emerged on Earth (>3.8Ga), but before the rise of oxygen in the Earth's atmosphere made $O_2$ a remotely detectable biosignature (Reinhard *et al*., 2017).

What we do know is that biological innovations on Earth have driven major changes in the redox state of our planet, leading to distinct observable states and planetary biosignatures (Kaltenegger *et al*. 2007). The prime example is the dramatic global scale transition to an oxidizing atmosphere resulting from oxygenic photosynthesis carried out by ancient cyanobacteria, detectable in the fossil rock record in the Great Oxidation Event (GOE) at 2.3-2.4 Ga (Luo *et al*., 2016). Innovations like this are dependent on the environmental conditions that allow them to arise and in turn drive the environment, leading to successional innovations and planetary geochemical states that co-evolve in a history-dependent manner. These states may be stable locally or globally at different temporal scales. Because *life and the planet's redox state co-evolve*, a complication for building probability distributions for *P(data*|life) is that the probabilities are time-dependent in a manner that depends on the states (*e.g.,* the dynamics are state-dependent, regarded as a hallmark feature of life (Goldenfeld and Woese, 2011; Walker and Davies, 2013), that is the probabilities we must construct are necessarily conditional.

On Earth, the changes in oxygen content of the Earth's atmosphere through geologic time corresponded with biological innovations of oxygenesis, nitrogen fixation, eukaryotic cells, multi-cellularity, and the arrival of plants on land (Ward *et al.*, 2016; Gebauer *et al.*, 2017; Berner *et al.*, 2003). Each may have been made possible by the changing geochemistry of the planet, sometimes fostered by earlier life, and there are a number of hypotheses to support this idea. For example, the input of $H_2O_2$ to the oceans from the thawing of a Snowball Earth state has been proposed to serve as a transitional electron donor in the emergence of oxygenic photosynthesis (Liang *et al.*, 2006). Another hypothesis is that the scarcity of ammonium as a nutrient in the face of oxidation at the GOE may have necessitated the development of nitrogen fixation (Blank and Sanchez-Baracaldo, 2010). The availability of oxygen allowed by aerobic respiration drives increases in organism size and productivity (Catling *et al.*, 2005). Additionally, the formation of the ozone layer from atmospheric oxygen altered the spectral quality of surface irradiance, protecting it from ultra-violet radiation and allowing the emergence of advanced life on land, as well as altering the color balance of light for photosynthesis (Kiang *et al.*, 2007a). Many of the evolutionary developments that brought about these transformations remain enigmatic. Some of these innovations have arisen independently multiple times, while others appear unique, some with evidence of an evolutionary pathway, but others without a clear origin. The probability of these innovations is a separate term in our Bayesian framework, *P(life),* to be treated in more detail below in Section 6.2.

For other planets, surmising evolutionary path and geological epochs for life that are not Earth-like offers a rich challenge for interdisciplinary science. One question is what false positives (high values of *P(data|abiotic)*) and negatives (low values of *P(data|life)*) arise over time). Oxygen and water on a young planet orbiting a flaring M star could be a false positive, or ambiguous where the age of the star and time scale for life's evolution are unknown (Luger and Barnes 2015).  False negatives may result when biosignatures are not detectable until long after initial development of the producing organisms. This could occur when time is required to build-up a biogenic product in an atmosphere, and for the planet's climate and geochemistry to shift to a different equilibrium, as exemplified by the possible 2 billion years between the first emergence of oxygenic photosynthesis and its detection in the Great Oxidation Event (Cole *et al.,* 2016; Lyons *et al.,* 2014).  Astrobiologists must accept that they are unlikely to detect marginal biospheres that have little detectable impact on a planet (*e.g.,* where *P(data|life) << P(data|abiotic)*, even if we expect *P(life) >0)*. In many cases, we may be unable to detect biosignatures from earlier organisms that are subsequently suppressed by later organisms and evolving chemical and climate conditions, those that may exist only in obscure niches such as deep hydrothermal vents, or are relicts in refugia toward the end of a planet's life (O'Malley-James *et al.,* 2013).  Yet, these marginal biospheres might explain extant life, being its precursor, or relict planetary chemistry. They are also important in estimating the prior probability of life *P(life),* and in obtaining reasonable estimates of the distribution of life on other worlds.

## 5.3.2   *Calculating conditional probabilities in biological evolution from past biogeochemical states*

From the foregoing discussion, it is clear that our planet has gone through many different states as Earth and its living systems have co-evolved with one another over geological time scales. Extending our understanding of life through its history, our *N=1* sample permits the study of many distinct biophysicochemical modes that differ from the current configuration of the coupled biosphere, geosphere and atmosphere. Within the exoplanet biosignature community it is often noted that life represents at least two end-member examples of an inhabited planet; the early Earth (low or absent oxygen) and the modern Earth (high oxygen) (e.g., Lyons *et al*., 2014; see also Meadows *et al*., 2017, this issue). Expanding this idea, each time life has radiated into a new geochemical niche could be considered as an additional data point: each example provides new insights into how selective processes can yield new biochemical mechanisms for energy acquisition and generation of biomass.

To understand potential biosignatures and their likelihoods requires linking the history of different modes of biological innovation and environmental states on Earth to their potential planetary-scale signatures. By extension, it is only by piecing together the histories of key molecular components that couple metabolic activity to planetary reservoirs that we may begin to estimate the temporal frequency and distribution of comparable biosignatures on other planets (Lyons 2014, Catling 2011). Two datasets, the geologic record and the genetic content of extant organisms, provide complementary insights into this history of how key molecular components have shaped or driven global environmental and macroevolutionary trends (Caron 2017; Kacar *et al*., 2017; Fisher 2016). Changes in global physiochemical modes over time are thought to be a constant rather than ephemeral feature, as life has continuously evolved protein functions for the >3.8 billion years of life's history on Earth. Organismal survival depends on how well critical genetic and metabolic components can adapt to their environments, necessitating an ability to adapt changing conditions. These adaptations can produce viable biosignatures where biological rates exceed abiotic ones, e.g. where *P(data|life)* > *P(data|abiotic)*.

The geologic record provides a number of biologically dependent indicators of macroscale atmospheric and oceanic composition, but provides little information by way of the exact behavior of the molecular components that altered the compositions of these reservoirs. One proposed way to infer the activity of ancestral molecular components is to reconstruct protein sequences that might have been present in ancient organisms, downselecting to a subset of possible sequences that may have been adapted to these ancient environmental conditions. It should be noted that such sequences are inferred based on the most parsimonious ancient sequence(s), given the diversity of modern ones, and are subject to historical ambiguity (Benner *et al.,* 2007; Kacar and Gaucher, 2012). With that caveat, reconstructing ancestral phenotypes that can lead to large-scale planetary biosignatures can be accomplished by identifying primitive biomolecular protein sequences that have impacted the cycling of C, N, S, O or P through global reservoirs (Kacar *et al.,* 2017). Studying the interface of past molecular behavior and environmental conditions may provide new insights into the interpretation of deep time biosignatures on Earth. For example, organismal and community fitness can be studied in the lab, as well as rates of production of biosignature gases. Data reconstructed through these studies may then be compared and contrasted with independent data obtained from primitive organic material for which there is a suspected or known biological imprint. Such findings may be critical for establishing biosignature baselines that are persistent and thus considered more broadly 'universal', or to identify cases where protein activity was not uniform in Earth's past. However, experiments incorporating these methods also require careful design to rule out signals from other potential artifacts impacting biosignature assessment, such as sequence reconstructive biases or organismal responses to non-adapted substituted components. Properly accounting for these ambiguities, the data generated from these approaches can inform how likely a given biosignature signal is within the space of understood catalytic proteins, contributing to our understanding of *P(data|life)*.

There is an analogous approach searching for ancestral metabolic pathways by comparing modern pathways, and in turn reconstructing putative ancient ancestors. This has been an established approach to studying the evolution of metabolic capabilities for over 70 years (see *e.g.*, Waley, 1969). Smith and Morowitz argue this approach can probe the pre-genetic epoch of biogenesis, and be used to understand the chemistry from which life arose (Smith and Morowitz, 2016). While our ability to reconstruct the exact history of life-on-Earth is debated, these approaches illustrate that a combination of genetic, structural and chemical analyses of the product of 3.5 billion years of evolution can provide insights into intermediate states in that evolution.

Understanding the function of ancestral components can also provide a novel means of gaining access to configurations of life that deviate from extant norms. For example, it is possible that life could have started with, evolved from, or subsisted by uptake of a different set of monomers than those utilized by current life on Earth (Forterre, 2007; Braakman, 2012). By engineering modern organisms with the behavioral properties of these ancient components, we may explore variations of so-called "weird life," which can yield yet more significant insights into the essential requirements for life as a universal phenomenon. This is a common approach in the field of synthetic biology, and indeed components of organismal genetic machinery have been successfully replaced with synthetic parts in functioning organisms. Examples include expanded genetic alphabets (Malyshev *et al.,* 2014), synthetic minimal bacterial genomes (Hutchison *et al., 2016*) ancient genes inside modern bacterial genomes (Kacar *et al., 2017*) and alternative nucleic acids (XNAs) (Taylor *et al.,* 2015).

## 5.4   Insights from Universal biology

All of the candidate biosignatures discussed thus far in this review have focused on chemical signatures of life. Searching for "life as we know it" implies searching for "biochemistry as we know it," *e.g.,* DNA, proteins and metabolisms like Earth's, such as oxygenic photosynthesis. Therefore, moving beyond "life as we know it" to "life as we don't know it" with unknown biochemistry will require developing new frameworks that address universal aspects of living processes. The idea of *"universal biology"* has been proposed with the intent of transcending the chemistry of life as we know it to uncover universal organizational properties of living systems (Goldenfeld and Woese, 2011, Davies and Walker, 2016) – perhaps associated with patterns in information flow or energy transfer – that should apply to any kind of life, even if it is based on a radically different biochemistry. Of the candidates for universal biology, chemical disequilibria has been the most widely discussed as a potential biosignature. But, it is unclear if disequilibria associated with life quantitatively differ from other planetary disequilibria, as discussed above. Additionally, some have argued that life exists to facilitate a more rapid approach to equilibrium than would be possible with geochemical processes alone (Shock and Boyd, 2015), such that living planets should be *closer* to equilibrium rather than farther as compared to non-living planets. Other candidates include universal scaling laws (West *et al.,* 2002; Okie, 2012), collective behavior (Goldenfeld and Woese, 2011), network structure (Jeong *et al.,* 2000), or informational structure (Davies and Walker, 2016). Although we are only in the early stages of developing a universal biology, insights into common organization properties of biological systems gained over the last decade hold promise for providing novel approaches to biosignatures in near-term searches for life and for longer-term mission planning, providing new frameworks for constraining *P(data|life)*.

### 5.4.1  Network biosignatures

Networks are used to quantify the properties of living systems across all scales of organization, from the chemistry within cells (Jeong *et al.,* 2000) and the structure of food webs (Dunne *et al.,* 2002), to the organization of cities (Bettencourt, 2013). A network is most simply described as the pattern of connections among a system of interacting entities. Mathematically, networks are studied using the tools of graph theory, where entities are represented by nodes and their interactions by edges. Familiar examples include social networks, such as Facebook, where individuals are represented by nodes and their friendships by edges (*e.g.,* an edge is present if two individuals "like" each other). Likewise, chemical species reacting with one another in a planetary atmosphere can be represented graphically with a network, where one node type represents molecular species and a second node type represents the reactions that occur among these species (left panel, Figure 7). Other graphical representations are possible, such as ones involving only molecular species (and no reaction nodes), which are connected if they participate in the same reaction (right panel, Figure 7).

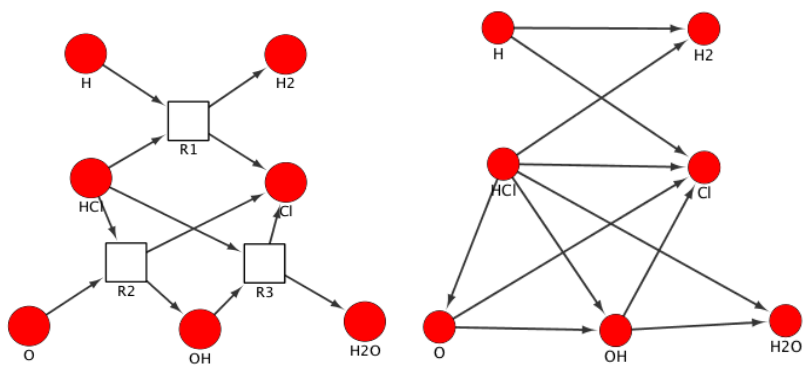

**Figure 7:** Two different graph-theoretic representations of the same chemical network, consisting of the reactions: H + HCl → $H_2$ + Cl, HCl + O → Cl + OH, and HCl + OH → Cl + $H_2O$. Network examples adopted from Sole and Munteanu (Sole and Munteanu, 2004).

Network representations have been used to study biochemical networks associated with metabolism. Jeong *et al*. (2000) demonstrated that the metabolic networks of 43 organisms, representing all three domains of life, are scale-free networks, meaning their degree distributions follow a power law $P(k) \sim k^{-\alpha}$, where $P(k)$ is the probability that a given molecular species participates in $k$ reactions (in network parlance $k$ is the degree of the node, corresponding to the number of edges connected to that node). Earth's metabolic networks are therefore highly heterogeneous in that there exist a few highly connected nodes (hubs) that link numerous less connected nodes together. This property has been explained in terms of enhanced robustness: heterogeneous networks are known to be more robust to the loss of random nodes than random networks. It is therefore a candidate signature of evolutionary processes at work, providing new ways to potentially constrain the value of $P(data|life)$.

The universality of metabolic network organization suggests that life on other worlds might evolve to exhibit similar network topology to that of Earth's metabolic networks, and therefore that network topology is itself a biosignature. One hypothesis is that life could additionally leave a topological imprint on atmospheric chemistry. To test this hypothesis, several studies have examined the network topology of Earth's atmosphere (Figure 8) and compared it to that of other worlds in our solar system (Gleiss *et al*., 2001; Solé and Munteanu, 2004; Holme *et al*., 2011; Estrada, 2012). The results of these comparative analyses indicate the Earth's atmospheric reaction network structure differs from other planetary atmospheres, and specifically that it is more like biochemical metabolism in its topological structure than it is like other atmospheres. In particular, Solé and Munteanu showed that Earth's atmospheric chemical reaction network exhibits scale-free topology, much like biochemical networks (Solé and Munteanu, 2004), whereas other planetary atmospheres, including Mars, Venus, Titan, the Jovian planets, are structured more like random networks. These results resonate with the view that life is indeed a planetary process and is deeply embedded in the Earth system, to the point that even the network arising from the chemical dynamics of the atmosphere is driven by life (and not just its molecular constituents such as $O_2$).

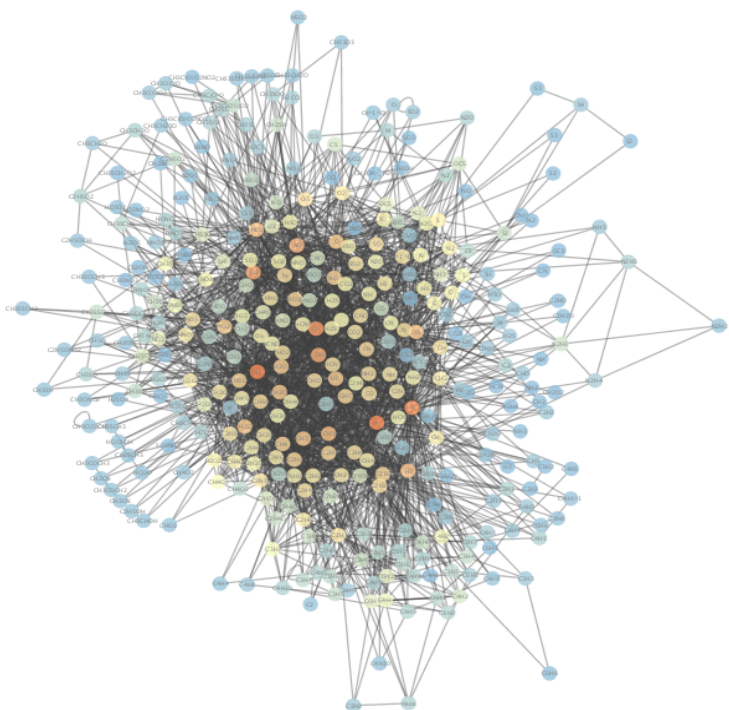

**Figure 8:** A network representation of Earth's stratospheric chemical reaction network. High degree nodes are highlighted in warm tones, and lower degree nodes in blue. Data from DeMore *et al*. (1997).

There are observational biases that must be accounted for in network analyses, as we know Earth's chemical constituents and its reaction network to a much greater level of detail than we do other planetary atmospheres. However, even the major constituents contained in Earth's atmosphere may require a more complex network to fully explain them. The fundamental reason for this is the same as the foundations of traditional thinking on "non-equilibrium" or non-steady-state biosignatures that has been in the minds of the exoplanet community for decades (Lovelock, 1965; see also Section 4.3). The introduction of non-steady-state gases by biology leads to additional atmospheric reactions that would otherwise not take place. Conversely, the presence of these same gases in a steady-state condition that does not require biological fluxes to maintain them implies that the chemistry of the atmosphere is such that their destruction rates are slow, a result of the atmospheric chemistry being less complex. We can consider the classic example of $O_2$ to illuminate this. The known mechanisms for accumulating detectable amounts of $O_2$ in a planetary atmosphere are all associated with atmospheres that are deficient in H. The results of this are atmospheres with chemical networks dramatically less complex due to the lack of H-bearing species and their reactions. Similarly, a planet without biological $O_2$ fluxes would not have the additional reactions that are caused by its presence in the atmosphere. The network complexity would instead be greatest when $O_2$ is present in an atmosphere that would otherwise destroy it rapidly. Similar trends are hinted at in studies of alternate biosignatures as well, such as the additional chemistry resulting from biogenic sulfur gases that causes $C_2H_6$ to be detectable in exoplanet environments (Domagal-Goldman *et al.*, 2011).

To validate this proposal, and utilize network-theoretic biosignatures for remote detection, several lines of research must come together. The properties that are unique to inhabited worlds need to be fully explicated. Recent work has shown that scale free topology is not as common as previously claimed (Clauset *et al.,* 2009), and requires rigorous statistical tools to confirm. In particular, relatively few molecular species are confirmed to be present in many planetary atmospheres, meaning atmospheric networks are small making it difficult to obtain statistically rigorous fits for the degree distribution. For exoplanets we will have even less data. Other topological properties must therefore be studied to determine in what ways Earth's atmosphere differs from other worlds. Topological properties should be analyzed across a range of kinetic and dynamic models, varying T, P and composition to determine how physical effects influence atmospheric reaction network topology to isolate possible biological origin, and constrain *P(data|life)* for network topology. A systematic analysis of the topology of different models for planetary atmospheres could be used to determine the likelihood of specific features, given planetary and stellar context. Additionally, deeper analysis should be done for Earth to determine how biology is driving the distinctive topological properties observed. Finally, if validated as a biosignature, it remains to be demonstrated how we can extract large-scale statistical properties of an atmosphere's network from the limited data we will obtain through remote observation. One possibility is to use Bayesian retrieval methods, used for extracting cloud properties from atmospheric data (Line *et al.*, 2012).

This concept is relatively new - to the exoplanet field at least - and thus warrants further investigation. Studies of the kinetic properties of various terrestrial worlds can test the overarching hypothesis that Earth's network is more complex, and that the increased complexity is due to biology. The application of this approach to multiple inhabited planets - including that of Early Earth - should also be conducted. This will allow us to understand how well this hypothesis holds, how useful it is to constraining *P(data|life)*, and in turn how useful it will be to future exoplanet astrobiology missions.

This illustrates one example of how considering the general, chemistry-independent properties of life may lead to specific research proposals in life detection strategies. Deeper research questions should also examine whether there are other, equally fundamental properties of life that can potentially lead to remotely detectable consequences.

## 5.4.2 Universal scaling laws, applicable to other worlds?

Another candidate for universal biology is the scaling laws associated with trends across different biological organisms. Familiar examples from physics include critical phenomena near phase transitions, where physical properties such as heat capacity, correlation length, and susceptibility all follow power-law behavior. Scaling relationships take the form $Y(\lambda N) = \lambda^{\beta} Y(N)$, where $\lambda$ is an arbitrary scaling parameter with scaling coefficient $\beta$, $N$ is typically a measure of the size of the system, and $Y$ measures a property characteristic of the system. Thus the scaling relation provides a direct mapping from the value of the parameter of interest, $Y$, for a system of size $N$, to the value of the same parameter measured on a system of size $\lambda N$. The scaling $Y(\lambda N)/Y(N)$ is then parameterized by a single dimensionless number, the scaling exponent $\beta$. A simple solution is the power law relationship associated with scale-free network topology as discussed in the previous section. In addition to power laws in networks, scaling relations have been studied in biology in phenomena as varied as patterns in species diversity (Locey *et al.,* 2016), the organization of cities (Bettencourt, 2013), and the structure of neural systems (Zhang and Sejnowski, 2000). A scaling relationship of interest for identifying universal patterns in biology, applicable to other worlds, are the allometric scaling relations, which relate features such as metabolic rate to body to size (West *et. al.,* 2002; Okie, 2012). These scaling relations change through the major transitions in biological architecture (e.g, from prokaryotes to protists to metazoans, see Figure 9) and appear to be universally held across life on Earth (Delong *et al.,* 2010).

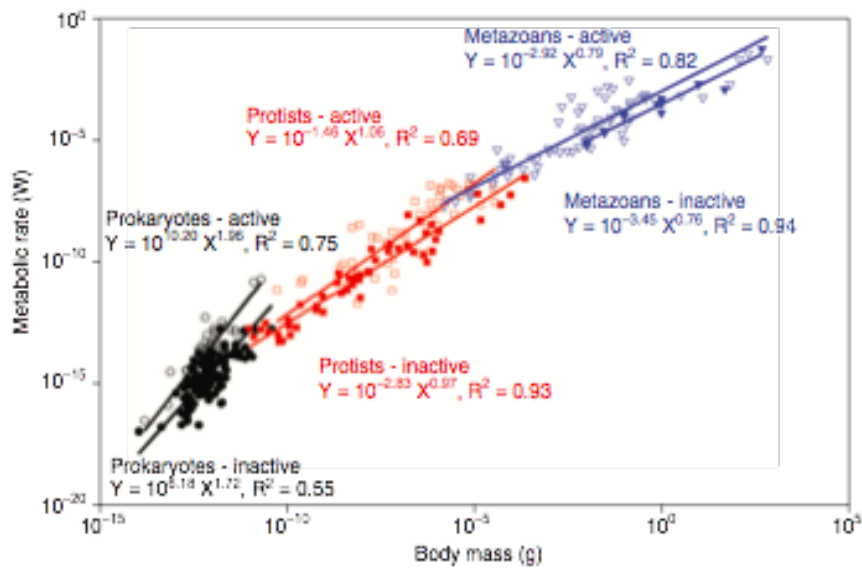

**Figure 9:** Empirically bserved scaling laws for metabolic rate as a function of body-mass exhibits three major regimes, associated with prokaryotes, protists and metazoans (Delong *et al.*, 2010). If these trends are universal and can be derived from an underlying common theory, it may be possible to apply the universal scaling relations to inform P(data|life) on other worlds. Figure adopted from Okie (2012).

Scaling relations, due to their ability to "predict" the values of system parameters based on other measured quantities, represent one of the closest approaches so far to a predictive theoretical biology, akin to theoretical physics. Using the observation that cells and organisms are constrained in their growth by resource distribution networks, predictive models can be generated that accurately provide values for the scaling exponents observed in a number of diverse biological systems (West *et al.,* 1999). However, there is as-yet no unified theory that explains the observed allometric scaling relations across different level of organization, nor when transitions in scaling regimes should occur. Nonetheless, the existence of these scaling relations suggests integrative frameworks for constraining *P(data|life)*. For example, modeling flux rates of biosignature gases on exoplanets could be informed by fundamental bounds on flux rates for

given biomass estimates provided by universal scaling (given minimal assumptions of a particular biological architecture), allowing us to extrapolate to metabolisms that might exist in non-Earth like environments. It should be noted that the universal properties associated with scaling laws and network structure are both *statistical* constraints on *P(data|life),* providing fundamental chemistry-independent bounds on our expectations for life to generate a given observable.

## 6   P(life)

So far, we have focused discussion on calculation of *P(data|abiotic)* and *P(data|life)*. The final term necessary for calculating the posterior likelihood of life (apart from knowledge of experimental noise) is the prior probability for life to exist in the first place, *P(life)*. This is the least constrained and most challenging term to quantify.  As we pointed out above, it is not sufficient to simply assign a probability that life-as-we-know-it exists on another world (which is unknown), but instead *P(life)* should be considered as decomposable into a family of conditional probabilities for the existence of different living processes on other worlds. Life is a path-dependent process, so each new biological innovation is dependent on those that preceded it, thus *P(life)* takes the form of conditional probabilities for each evolutionary step (which itself may be difficult to define, what are the relevant steps?). As an example, in assessing the probability for multicellularity to evolve, we might decompose it into the following series of conditional probabilities:

*P(multicellularity) = P(multicellularity|eukaryogenesis)P(eukaryogenesis|emerge)P(emerge)*

where we have here considered only two of the most significant "major" steps. There are many more unspecified steps that will be necessary to articulate to map the limited observables of exoplanets to a reasonable estimate of the prior probabilities of living processes. The prior probability of every living process will ultimately depend on *P(emerge)*, the probability for life to originate. Thus, how well we can constrain *P(life)* depends on how well we can constrain the probabilities for the candidate living processes that may have generated the signal, its evolutionary history, and ultimately the origins of life.

## 6.1   *P(emerge)*: Constraining the probability of the origins of life

As noted in Section 3 introducing the Bayesian approach, attempts have been made to constrain *P(emerge)* within a Bayesian framework by Carter and McCrea (Carter and McCrea, 1983) and more formally by Spiegel and Turner (Spiegel and Turner, 2012), with the conclusion that *P(emerge)* could be arbitrarily close to 1 or zero. In other words, *P(emerge)* is currently unconstrained (apart from the trivial statement that it is not identically zero). In Spiegel and Turner (Speigel and Turner, 2012), it was estimated that the likelihood for the emergence of life follows a Poisson distribution, such that life was most likely to arise early in a planet's evolution. However, for M dwarf stars the early stellar environment may not be conducive to life (see Section 4.1 on stellar context) and in general it is unknown if life must arise early in a planet's evolution, or whether it could occur at any time. Ideally, we would be able to calculate *P(emerge)* from theory, but there currently are no theoretical bounds -- we do not have a quantitative definition for life nor a theory of the emergence of life from which to do such *ab initio calculations* (Walker, 2017).  We must better understand the mechanisms underlying the origins of life to make a case for *P(emerge)*.

If the emergence and evolution of life requires time, then knowing the age of the star is valuable to assessing the potential for life, and thus avoiding systems where *P(life)* might be too low to confidently detect life. Measuring the ages of stars is more accurately done for the youngest ages when the youth diagnostics (such as spectroscopy, photometry, kinematics, chemistry, etc.) are more clearly measureable.  For stars much older than a billion years we must rely on less precise techniques.  However, from statistical studies of many stars in various regions of our galaxy, we are able to better refine the ages of old stars by their location in the galaxy (*i.e.,* disk or halo), by the composition of the stars (high or low metallicity) and by their level of stellar activity as most stars tend to be less active (*i.e.*, less flaring, slower rotation) as they age.

In Scharf and Cronin (2016) a formalism akin to the Drake equation was proposed for estimating *P(emerge)* at a planetary scale. The mean expected number of abiogenesis events on a planet in a given time interval was suggested to depend on four parameters: the number of potential building blocks, the mean number of building blocks per organism (acknowledging ambiguities in the definition of "organism"), the size of the subset of building blocks available to life during a fixed time interval, and the probability of assembly of those building blocks.  The latter term, based on the probability of assembly, was intended as a "catch-all" that does not require detailed knowledge of mechanism and could, for example, include the probability per unit time of vesicle self-assembly, or a sequential series of steps leading to an evolvable system. In this formulation, this probability is the least constrained parameter, necessitating input to constrain it from *in vitro* and *in silico* research.

One potentially fruitful path for estimating these probabilities is "messy chemistry", where the goal is to study the statistical properties of chemical systems and their interactions with other compounds, formation structures, *etc.* in cases where precise composition and mechanism are not known (Guttenburg *et al.,* 2017). Here "messy" refers to high diversity of products, intermediates, and reaction pathways that cannot all be precisely identified. Analysis of the bulk properties of geochemical organic samples provides one example, where the Petroleum industry has developed methods for classifying crude oils quickly by measuring a small subset of its properties.  Applied to chemistries relevant to the origins of life, the emergence of 'life-like' features could be studied statistically to estimate the likelihood of functional polymers emerging from random mixtures, as just one example. This could in turn be tied to different environmental contexts through explorations of a diversity of environmental parameters. One such parameter is the rate of hydration-dehydration cycling, where wet-dry cycles have recently become a prominent mechanism in origins of life research for driving the abiotic synthesis of far-from-equilibrium biopolymers (Hud and Anet, 2000; Walker *et al.,* 2012; Mamajanov, *et al*., 2013). For polypeptide synthesis, the duration of the dry-phase has been shown to affect product yield, along with other environmental parameters such as temperature, number of cycles, initial monomer concentrations and pH (Rodriguez-Garcia *et al*., 2015). For cycling driven by tides, or day-night cycling, this type of data could place bounds on *P(emerge)* for exoplanets based on their rotation rate. These and other environmental parameters should be explored in large-scale, parallel chemical experiments to generate the necessary statistics.

In Scharf and Conin (2016), it was also suggested that multiple planet systems would have a higher value of *P(emerge),* because impact ejecta exchanged between neighboring planets with parallel chemistry and parallel chemical evolution could enhance rates for development of molecular complexity. Future modeling will need to determine if rates for the origins of life (or its persistence and evolution) are enhanced for worlds that are closely neighboring other habitable worlds or not.

## 6.2    Biological innovations and the conditional probabilities for living processes

Since the emergence of life on Earth, life has undergone a number of different stages of evolution (Szathmary and Maynard Smith, 1995, Braakman and Smith, 2012; Bains and Schulze-Makuch, 2016). Tracing the history of biological innovations allows the possibility of leveraging the diverse history of life on Earth, where $N=many$ (see Section 5.3), to infer the likelihood of evolutionary events based on their frequency of independent origins. However, care must be taken to determine what is meant by 'independent'. For example, the probability of evolving multicellularity is dependent on the prior probability of eukaryogenesis (at least for Earth life), since all multicellular organisms are eukaryotes (as in the example above). It is also dependent on the prior probability of photosynthesis, since complex life evolved in the presence of $O_2$, and because $O_2$ is widely believed to be a precondition for large, multicellular animal life. There are many more such conditional probabilities we could assess. Thus while the transition to multicellularity is itself a common occurrence in the history of life on Earth, multicellularity may not be universally common on inhabited planets if either the probability of eukaryogenesis or oxygenic photosynthesis, or any other steps in the pathway to multicellular life are rare. We also must consider that all life on Earth shares a common ancestry, so when considerations of evolutionary 'independence' get blurrier the further we trace conditional probabilities for evolutionary events into deep history. Until we discover another example of life with independent origins (or have a guiding, universal theory), it will be difficult to say with certainty what the likelihood of similar evolutionary events would be from a different starting point (e.g., a different biogenesis, with different chemistry).

Rare events in evolution are often associated with major transitions or innovations. Maynard Smith and Szathmary identified eight major transitions, with respect to changes in units of selection, in the history of life on Earth, each associated with transitions in the nature of information transfer between and within individuals (Maynard Smith and Szathmary, 1995). These include the transitions of: replicating molecules to populations of molecules in compartments, unlinked replicators to chromosomes, the RNA to DNA-protein world (genetic code), prokaryotes to eukaryotes, asexual clones to sexual populations, multicellularity, eusociality and linguistic societies. Missing from Maynard and Smith's scheme are metabolic innovations that did not necessarily change what constitutes a selectable individual, but nonetheless had a significant impact on the biogeochemical evolution of the Earth-system, such as the origins of photosynthesis and nitrogen fixation. An important question for understanding the evolution of life on Earth is whether or not we should expect the same innovations to occur again if we "rewound the tape of life" (Gould, 1989). A more critical question for exoplanet biosignature research is: how frequently should we expect these same, or different biological innovations to happen on other inhabited worlds?

In Bains and Shulze-Makuch (2016), three possible hypotheses were proposed for the conditional probabilities of evolutionary innovations to occur: the critical path hypothesis, the random walk hypothesis, and the many paths hypothesis. In the critical path scheme, innovations require preconditions that take time to develop (determined by the nature of the event and the geological and environmental conditions of the planet), but once the preconditions exist the event happens on a well-defined timescale. In the random walk hypothesis, the innovation is unlikely to occur (being based on one or more highly improbable events) thus significant time must elapse (on average) before the event occurs (a complication arises in interpreting this hypothesis in the event of post-selection, where the event already occurred, in which case a rare event could have happened rapidly, see Carter (2008) for discussion). In the many paths hypothesis, the innovation requires many random events to create a complex new function, but many combinations can generate the same functional output, so the chance of the innovation is high. The key steps in the evolution of life, development of prebiotic chemistry, synthesis of cellularity, and the invention of metabolism and what pathway they took is not entirely clear. While modern geological, chemical, and genomic methodological approaches are extremely powerful, there nevertheless remain considerable challenges to precise understanding of the events leading to the expansion of life on our planet. Considerable effort may be required to understand the detailed history of evolution from methods as varied as radiocarbon dating, paleobiology, and molecular evolution.

We summarize here some critical gaps in our knowledge of early events in the evolution of life on Earth. The formation of a cell envelope serving as a permeability barrier and preventing the free diffusion of chemicals into and out of cells and ability to produce and store cellular energy was probably an essential prerequisite of the emergence of life on Earth. It is unclear if life is necessarily 'cellular', although it is difficult to envision possible alternatives. Our current knowledge of how chemiosmotic coupling and cellular energy metabolism evolved from cellularity and set the stage for evolution of the last universal common ancestor (LUCA), most likely with an universal genetic code already established, which subsequently evolved into all life on earth is sketchy at best (Weiss *et al.*, 2016).

The very early appearance of the electron transport chain, membrane complexes responsible for fundamental respiratory processes in all prokaryotic cells, also remains an enigma. While they are found in both Bacteria and Archaea on Earth and may be used for redox reactions of nitrogen, sulfur, and oxygen gases, the precise evolutionary steps taken in their development still confound modern molecular phylogenomic analysis (Castresana and Saraste, 1995). Similarities between a variety of chromophores – e.g. porphyrins with different metal ion centers within membrane proteins of the electron transport chain and photosynthesis – remain tantalizing (see review in Schwieterman *et al.*, 2017, this issue). Generation of an electrochemical gradient used to drive ATP synthesis via chemiosmotic coupling is a common nearly universal theme (Racker and Stoeckenius, 1974), the basis of which deserves further attention.

The emergence of oxygenic photosynthesis involved conservation of photosystem structures that evolved in earlier organisms but also added a unique molecule, the oxygen evolving complex (OEC), which is responsible for oxidation of water but whose origin remains elusive. A number of hypotheses have been forth, though none yet has reached consensus. . These hypotheses span approaches from the biophysics of transitional electron donors, generally in the context of geochemical environment (Blankenship & Hartman, 1998; Dismukes et al., 2001; Sauer & Yachandra, 2002; reviewed at the time by Blankenship *et al*., 2007; Fischer *et al*, 2016); phylogenetics to constrain lineage and timing (Xiong and Bauer, 2002; Soo *et al.*, 2017); and reconstruction of evolutionary relationships among the reaction center proteins and their biosynthesis pathway (Cardona *et al.*, 2015; Cardona, 2016). This work to date advances knowledge of the origins of oxygenic photosynthesis on Earth. However, the likelihood of the OEC capability arising given a conducive geochemical setting, that is, the likelihood of another planet developing oxygenic photosynthesis by the same method is difficult to constrain.

Biological innovations may be classified through disciplinary perspectives in addition to that of biologists, raising the question of where to draw the black box for biosignatures. For example, light-capturing chemistry is not unique to photosynthesis, but a range of other structurally and evolutionarily unrelated pigments are used for the capture of light energy (DasSarma, 2006). The key pigment in terrestrial photosynthesis – the family of chlorophyll molecules – is structurally related to a number of other porphyrin derivatives such as haeme used in energy transfer and oxygen handling (see summary in Schwieterman *et al*., 2017, this issue). However, other light-capturing pigments include a range of isoprene pathway derivatives such as carotenoids and retinal, although not all of these are used to for obtaining electrons for carbon reduction on Earth. This both suggests that the evolution of light-capturing chemistry is not a unique event and opens questions about what features are universal (as opposed to evolutionarily contingent) that we might search for signs of on other worlds. At the same time, whether some types of pigments had to evolve first or were completely independent, and their evolution, relative abundance, and byproducts are of significant interest for detection of biosignatures.

# 7   A Bayesian Framework Example: Surface Biosignatures

We next provide an illustrative example of how one might apply the Bayesian framework,

considering a hypothetical case where the observational data exhibit a surface spectral feature, $S_{obs}$. Let's assume $S_{obs}$ indicates an observed surface spectral signal that appears anomalous compared to *P(data|abiotic)*, while gaseous signals are ambiguous about the possibility of life. Therefore, we wish to know if the surface feature is a sign of life. Our general equation is (excluding noise for simplification in Eq. 2):

$$( 10 )\; P(life|S_{obs}) = \frac{P(S_{obs}.|life)\; P(l\;fe)}{P(S_{obs}.|abiotic)(1-P(life)) + P(S_{obs}.|life)\,P(life)}$$

We next briefly outline the data, modeling and challenges to overcome to gain quantitative understanding of each of the terms in the right-hand side of Eq. 10 for the surface biosignature example (other biosignatures would likewise require a convergence of better data and models to realize the goal of quantifying the likelihood of life in a given observational signal).

**$P(S_{obs}|life)$, the likelihood of the observed data given life is present.** As reviewed in Schwieterman *et al.* (2017, this issue), "edge spectra" are characteristic of the reflectance of many basic biomolecules (Poch et al. 2017) and a wide variety of known pigmented organisms that exhibit spectral features throughout the visible and near-infrared range, including oxygenic and anoxygenic phototrophs, as well as non-photosynthetic organisms (Hegde *et al.*, 2015; Kiang *et al.*, 2007b; Schwieterman *et al.*, 2015). The best known "edge" feature and the only one clearly vetted as an unambiguous biosignature on Earth is the "vegetation red edge" (VRE): the steep increase in reflectance between red and near-infrared wavelengths in plant leaves (Des Marais *et al.*, 2002; Gates *et al.*, 1965; Sagan *et al.*, 1993; Seager *et al.*, 2005; Tucker, 1979). Pigments whose function involves selective interaction with radiation at particular wavelengths, such as for light harvesting or photoprotection, are partly governed by efficiencies that can help to constrain their likely spectral expression (Larkum and Kuhl, 2005; Stomp *et al.,* 2004; Stomp *et al.,* 2007a). But a confident predictive capability still requires additional research on evolutionary path and efficiency limits (Li and Chen 2015; Marosvölgyi and Gorkom 2010; Milo 2009; Punnoose *et al.,* 2012). Other pigments can have colors that are fortuitous independent of their primary function in the organism, such as with carotenoids in haloarchaea (Schwieterman *et al.*, 2015), and the community has barely begun to vet these as biosignatures. In general, all pigments can have variable spectra tuned within different pigment-protein complexes. Not just one surface biosignature may present itself, but a combination may be present, such that $P(L_n) > 0$ for more than one living process, as observed by Parenteau *et al.* (2015) in the reflectance of microbial mat communities. Thus, science is still at a beginning stage for quantifying ***P(S_{obs}|life)***, which must be expanded into the sum of probabilities for a large number of possible life processes that can generate surface spectral signatures that accounts for acclimation and adaptation to the local environment.

***P($S_{obs}$|abiotic), the likelihood of the observed data given life is not present.*** There are known abiotic 'edge'-producing minerals, including elemental sulfur and cinnabar (Seager *et al.*, 2005). To rule out false positives or 'edge spectra' as biosignatures, there is need for a systematic compilation and classification of 'edge' wavelengths in minerals and other abiotic materials. Already, 10,000+ reflectance spectra exist in disparate databases from multiple groups (e.g. Baldridge *et al.*, 2009; Clark *et al.*, 2007). A collation of these spectra, and consideration of their likelihood on a given planetary surface, could be used to construct probability distributions for the likelihood of an abiotic 'edge' at different wavelengths. A putative biosignature candidate could be compared with this distribution as one estimate of biogenicity. This approach could be

combined with a Bayesian framework applied to the evaluation of gaseous biosignatures (for example, see Catling *et al*. (2017, this issue)). It is important to note that this technique would require incorporation of a physical understanding of planetary surface processes, and a method for adjusting probabilities given planetary observables such as density, instellation, observed atmospheric composition, etc., while considering uncertainties in each variable.

The search for polarization signatures could also assist in fingerprinting a biogenic "edge" signature (Sparks *et al*., 2009a, 2009b; Berdyugina *et al*., 2016; Patty *et al*., 2017). Both linear and circular polarization measurements may aid in this effort with the tradeoff being that circular polarization signatures are more uniquely biological, but yield fainter signals, while linear polarization signatures are generally stronger, but more susceptible to false positives. Ultimately, the degree of polarization measured may be considered an input variable in a more robust, probabilistic framework for evaluating surface biosignatures.

$P(life)$ *the prior probability of life*.   The prior probability for living processes will depend on the particular process and the evolutionary paths permitting its emergence. An example is oxygenic photosynthesis. One of the greatest challenge in calculating *P(OP)*, the probability of oxygenic photosynthesis, is the unknown origins of the oxygen evolving complex (OEC), the key molecule in oxygenic photosynthesis.   *P(OP)* can be further expanded into conditional probabilities on the origin of other key processes and on the origin of life *P(emerge)*. The probability of all living processes ultimately depends on the prior probability for life to emerge in the first place.  For oxygenic photosynthesis, the probability of the observed state if oxygenic photosynthesis is or is not present will depend on several angles of understanding of what oxygenic photosynthesis is and does. For example, can its existence be surmised from black box net fluxes of gases, or is a process-based understanding required? And, at what stage it could appear in life's coevolution with the planet (what predecessors are necessary and likely)? As Meadows *et al.* (2017, this issue) cover in detail, we may see false positives or negatives for oxygenic photosynthesis. This example is by no means comprehensive of all the nuances of biophysics, biochemistry, biogeochemistry, ecophysiology, ecology, and evolutionary history that play a role in the emergence and expression of a surface signature of oxygenic photosynthesis. However, it highlights some of the many challenges the community will face in coming decades to constrain priors for the probabilities of life processes.

From the above examples of factors that will contribute to our ability to assess $P(S_{obs}|life)$ and $P(S_{obs}|abiotic)$ it is clear that many processes must be known to constrain these likelihood functions and to constraint *P(life)*. The conditional probabilities $P(S_{obs}|life)$ and $P(S_{obs}|abiotic)$ can be expanded using the law of total probability:

( 11a ) $P(S_{obs}|life) = \sum_n P(S_{obs}|life \cap L_n)P(L_n|life)$

( 11b ) $P(S_{obs}|abiotic) = \sum_n P(S_{obs}|\neg life \cap A_n)P(A_n|\neg life)$

where $L_n$ are all of the types of *living processes* which could, in principle, generate the surface signal, and likewise $A_n$, are all the abiotic processes that could generate the signal in the *absence* of life. Since $L_n \in life$, we can write $P(L_n) = P(L_n|life)$, and the above simplifies to $P(S_{obs}|life) = \sum_n P(S_{obs}|L_n)P(L_n)$ (likewise for the abiotic expansion where we use the fact that $A_n \in \neg life$). $S_{obs}$ could arise due to one of the processes $L_n$ or a combination (note any living process that does not generate the observed signal $S_{obs}$ will drop out of the sum in Eq. (11)), likewise for $A_n$. The processes $L_n$ and $A_n$ will in general be conditional probabilities, and depend on the likelihood function of a particular series of biological or planetary evolutionary events, respectively. Ultimately these will depend on the data available to us such that:

(12) $P(L_n) = f(Star, M, \rho, o, c)$ and $P(A_n) = f(M, \rho, o, c)$

where Star, M, ρ, o, and c are the planet's parent star characteristics, mass, density, orbital parameters, and expected elemental composition, respectively – that is, parameters which we can observe, but that we do not expect to depend on the presence of life.

Additionally, *P(life)* could decomposed into a sum of probabilities of all living processes for which we expect to have a nonzero prior on the planet of interest. Ultimately these will all depend on the prior for the origins of life on that world, so we here approximate $P(life) \cong P(emerge)$ (where $P(emerge)$ is also a function of Star, M, ρ, o, and c), which assumes that the largest evolutionary bottleneck is the origin of life itself. Given we do not know the prior probability for life, this seems as reasonable a starting assumption as any (*e.g.*, we cannot detect life if it has not emerged). A goal for future research must be to better constrain the priors for particular evolutionary events (or sequences of events) and for the origins of life (*e.g.,* the limiting event may not be the emergence of life, but the evolution of oxygenic photosynthesis or some other evolutionary outcome).

Our final equation, accounting for the signal, S, is:

( 13 ) $P(life|S_{obs}) = \dfrac{\sum_n P(S_{obs}|L_n)P(L_n)\,P(emerge)}{\sum_n P(S_{obs}|A_n)P(A_n)\,(1-P(emerge))+\sum_n P(S_{obs}|L_n)P(L_n)\,P(emerge)}$

For any putative evidence for life, the above will contain many terms with many conditional dependencies. It is a challenge for the community to synthesize many research areas to constrain the probabilities for observational signals from both abiotic and living systems. In Figure 10 we map out just a few constraints on the components of the above equation for the surface biosignature example according to current knowledge.

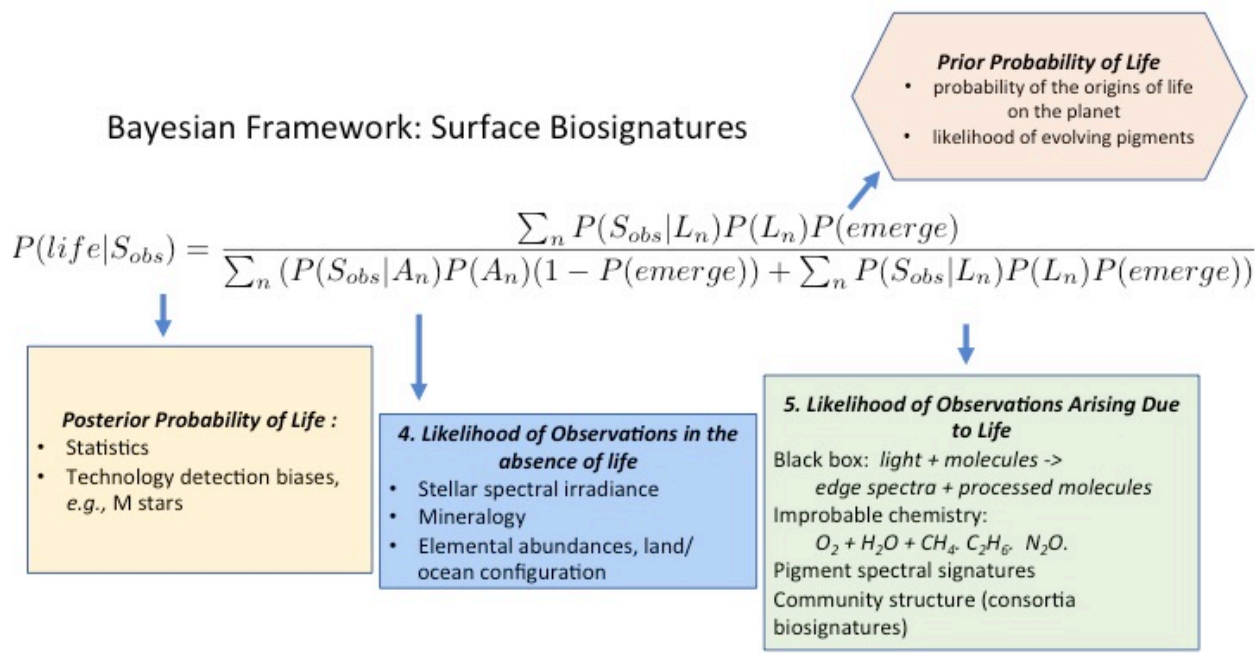

**Figure 10:** An example application of Bayesian framework for assessment of the likelihood of life due to the measurement of surface spectral signal, which is a putative biosignature.

# 8   Tuning search strategies based on the Bayesian Framework

The continued rapid increase in discovered planets in the coming years will make it necessary -- as well as possible -- to calculate the likelihood of a given observation being produced by an *inhabited* planet. This, in turn, requires a concerted effort to build comprehensive systems models of planets that include the myriad interactions of the biosphere with other planetary systems. Such models must also be flexible to be applicable to planets with a variety of compositions, sizes, orbital properties, orbiting stars with a variety of properties, etc. In this paper, we will have presented a discussion of the necessary tools, disciplines, and methodologies necessary to build, assess, and improve such models and improve our estimates of *P(data|life)*, *P(data|abiotic)* and *P(life)*.

Based on the statistical framework provided by the Bayesian approach, there are two strategies that can be employed in the development of future missions to search for life: the first is to maximize our confidence in *P(data|life)* and the second is to maximize our confidence in *P(data|abiotic)*. These are not necessarily mutually exclusive, and we provide two examples to provide insights into how our confidence in these terms, coupled with constraints on *P(life)*, should inform mission design.

The likelihood that a given observed signal is a product of life, *P(data|life)*, is best improved by observing a single target over long periods of time. We might expect that over time, observations of an inhabited planet would lead to an increase in *P(data|life)*. Conversely, subsequent observations of an uninhabited world would lead to a decrease in *P(data|life)*. As an example, we can consider the case of an inhabited planet, observed with a UV-Visible-Infrared mission such as HabEx or LUVOIR (see Fujii *et al.*, 2017, this issue). Over time, either of these missions would provide continued accumulation of knowledge of the exoplanet system. In terms of biosignatures, such a mission might first detect $O_3$, then $O_2$, then $CH_4$. This would be followed by tighter constraints on the compositions of each of these gases, and detection and measurement of the surface distribution of oceans, continents, and potentially a red edge-like effect from pigments. Eventually, we may see seasonal variations in gases (Fig. 11) produced or consumed by biology. And with subsequent missions operating at different wavelength ranges, the surface

climate could be determined, isotopic measurements made, and trace gases identified. In each of these steps, our confidence in the presence of life, *P(data|Life),* would increase.

One of the advantages of a well-developed Bayesian framework is that it can quantitatively inform observational strategies. For example, in the above scenario, Bayesian analysis would provide metrics to determine the more valuable observations for placing tighter constraints on the concentrations of biogenic gases, or observations of the temporal variability of those gases over seasonal time periods.

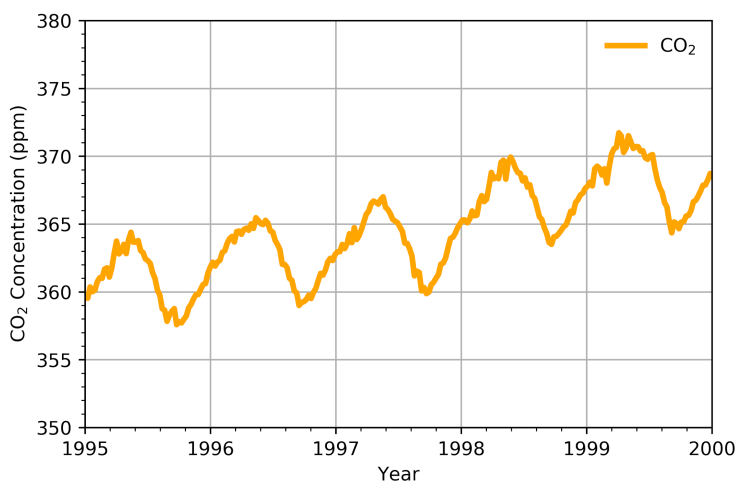

**Figure 11:** Seasonal variation in $pCO_2$ as an enhancement in P(data|life). Volume mixing ratio measurements $CO_2$ are sourced from the National Oceanic and Atmospheric Association (NOAA) at Mauna Loa, Hawaii, USA for the 1995-2000 time interval (Thoning *et al*., 2017). The seasonal change in $CO_2$ in the northern hemisphere is mostly reflective of the seasonal growth and decay/senescence of land-based vegetation (Keeling *et al*., 1996). These data were obtained from the NOAA's Earth System Research Laboratory (https://www.esrl.noaa.gov/).

Another scenario is that the observational spectra do not match our models, which could occur either for *P(data|life)* or for *P(data|abiotic)*. Although there are challenges with constraining both *P(data|abiotic)* and *P(data|life),* arguably the latter term is the one that will provide the greater statistical uncertainty. One question is then, how can we best constrain *P(data|abiotic)* based on our planetary models to maximize our confidence that deviations from the expected observations arise due to life?

The best way to constrain *P(data|abiotic)* will be to conduct large statistical surveys of uninhabited worlds, as discussed in Section 4. So far, data sets of this nature are scarce for exoplanets and non-existent for Earth-like worlds. One notable example for hot Jupiters is the 19 transiting examples with published transmission spectra obtained with the Hubble/WFC3 G141 near-IR grism. A majority of these (10 of 19) report a detection of $H_2O$ in their atmosphere (see list in Iyer *et al.,* 2016). Recently, it was shown that the individual spectral of these planets coherently average to produce a characteristic spectrum (Iyer *et al.,* 2016), which is reproducible with simple forward models, providing confidence that there exists a representative spectrum for at least a significant fraction of hot Jupiter exoplanets (Fig. 12). In this case, individual spectra do not fit our models well, due to stochasticity in planetary evolution, and in our measurements. However, the characteristic, averaged spectrum can be reproduced by models. It is an open question whether representative spectra will also accurately describe ensemble spectra of Earth-like worlds. Given the limited data we can collect on exoplanets (see review of observation capabilities in Fujii *et al*., 2017, this issue), and the stochasticity of planetary evolution (Lenardic, *et al*., 2012), it may be that we are only able to predict exoplanet spectra with high confidence probabilistically.

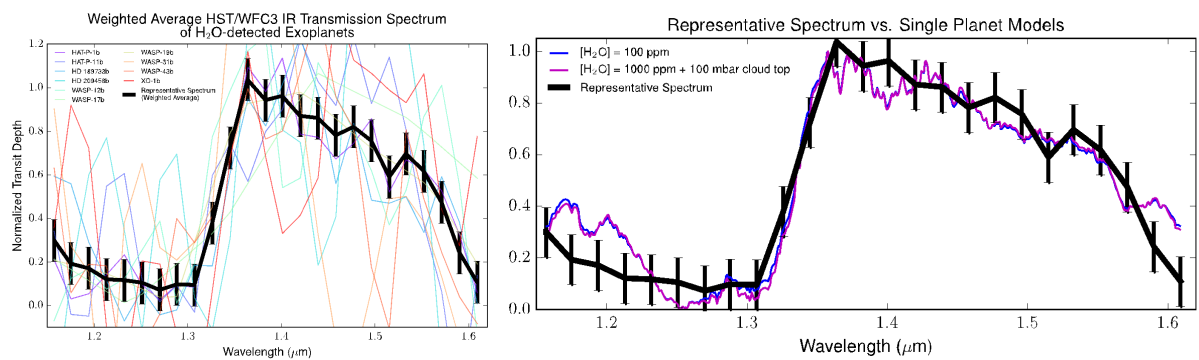

**Figure 12:** Left: Normalized HST/WFC3 IR transmission spectra of 10 exoplanets with reported H$_2$O-detections combined with a weighted mean to create a representative spectrum of H$_2$O-bearing exoplanets. Right: Comparison of representative spectrum (black) to single planet models (see Iyer *et al*., 2016 for details). Figure from Iyer *et al*. (2016). If this turns out to be the case, the community may need to shift focus to thinking about also detecting life *deterministically*, by analyzing coherently averaged spectra of many candidate worlds.

An advantage of a statistical approach to life detection is that it allows for the combination of a range of observations, including integrating over time and sampling large statistical data sets and can place bounds on the three terms of the Bayesian framework. For example, we can sum observations across planets and ask how our confidence that life exists on *an* exoplanet is changed by a new observation. If *P(life)* = 0.25 for an ensemble of 11 exoplanets, then a simplistic calculation suggests that we have a 95% confidence that life exists on at least one of them. Here a null result is almost as significant as a positive one. If we survey a sample of planets that are candidates for supporting a particular life process, and we find no evidence of that process on those worlds, we gain important information for constraining *P(life)*. This kind of analysis can also provide a guide to direct how many planets we must survey to detect life depending on how frequently we expect life to occur. The actual (as yet) unknown value of *P(emerge)* is critical to determining the most effective search strategy. If life is common (*P(life) >> 0*), it makes sense to target individual worlds and obtain high-resolution spectra, as is the proposed search strategy for JWST. However, if life is uncommon, we may be highly unlikely to be so lucky as to stumble on the right target. In this scenario, a more optimized search might, for example, take lower resolution spectra of more worlds to generate high-confidence representative spectra. Signatures of living processes might then be inferred from large statistical data sets of planet observations in cases where our expectations do not match averaged spectra (in cases where life is uncommon, but not rare). This has the advantage of enabling searching for "life-as-we-do-not-know-it", which is a challenge for current search strategies.

## 9    Conclusions

In this paper we set possible future directions for research on exoplanet biosignatures, highlighting promising directions that are not yet mainstream methods, but hold potential to revolutionize our search strategies in years to come. A major hurdle to be overcome in the coming decades is our lack of constraints on *P(life)*, and in particular *P(emerge),* the likelihood for the origins of life. This is one area where exoplanet science will need to interface with new communities, including those studying evolutionary biology, the co-evolution of Earth and life, and the origins of life. Exoplanet scientists will gain knowledge of constraints on the relevant terms in the Bayesian framework, informed by Earth's life and attempts to attract universal principles. In turn, our expanding observational searches for life should take advantage of the ensemble statistics we will be able to generate in the coming decades to inform our understanding of the distribution of life, placing additional observational constraints on *P(data|abiotic), P(data|life)* and possibly even *P(life),* for example, by identifying planetary

environments where no life is found (*P(data|abiotic)* matches well with observations) and those environments where life may be generating the observed spectra (eight *P(data|abiotic)* is not explanatory, or *P(data|life)* is). By combining the efforts of these diverse communities, combining deep knowledge of Earth and its life, with constraints afforded by the plurality of exoplanets, we have the first opportunity in history to put quantitative bounds on the distribution of life in the universe. As emphasized in this paper, this will require a concerted multidisciplinary and international effort, as should be expected from the enormity of the task of discovering life beyond Earth.

# 10 Acknowledgments

EWS is grateful for support from the NASA Postdoctoral Program, administered by the Universities Space Research Association. SD was supported by NASA Exobiology grant NNX15AM07G.

0